\documentclass[sigconf]{acmart}

\makeatletter
\let\@acmBadgeL@image\@empty
\let\@acmBadgeR@image\@empty
\makeatother

\setcopyright{none}
\renewcommand\footnotetextcopyrightpermission[1]{}
\setcopyright{none}
\makeatletter
\renewcommand\@formatdoi[1]{\ignorespaces}
\makeatother
\def\BibTeX{{\rm B\kern-.05em{\sc i\kern-.025em b}\kern-.08emT\kern-.1667em\lower.7ex\hbox{E}\kern-.125emX}}
    
\usepackage{nicefrac}
\usepackage{siunitx}
\usepackage{array,framed}

\usepackage[normalem]{ulem}   
\usepackage{xcolor}

\newif\ifshowdiff
\showdifftrue

\newcommand{\removecolor}{red}

\DeclareRobustCommand{\rm}[1]{%
  \ifshowdiff
    {\color{\removecolor}\sout{#1}}%
  \else
    \relax
  \fi
}

\usepackage{booktabs}
\usepackage{
  color,
  float,
  epsfig,
  wrapfig,
  graphics,
  graphicx,
  subcaption
}

\usepackage{textcomp}
\usepackage{amsmath}
\usepackage{setspace}
\usepackage{latexsym,fancyhdr,url}
\usepackage{xurl}
\usepackage{mathtools}

\usepackage{enumerate}
\usepackage[ruled, vlined, linesnumbered]{algorithm2e}
\usepackage{anyfontsize}

\usepackage{xparse} 
\usepackage{xspace}
\usepackage{multirow}
\usepackage{csvsimple}
\usepackage{balance}
\usepackage{enumitem}
\usepackage{amsthm}

\usepackage{pifont}
\newcommand{\xmark}{\ding{55}} 
\newcommand{\cmark}{\ding{51}} 
\colorlet{mygreen}{green!65!black}
\newcommand{\hl}[1]{\textcolor{black}{#1}}

\usepackage{
  tikz,
  pgfplots,
  pgfplotstable
}
\usepackage{hyperref}

\usetikzlibrary{
  shapes.geometric,
  arrows,
  external,
  pgfplots.groupplots,
  matrix
}

\usepackage{listings}

\lstdefinelanguage{Rust}{
  morekeywords={fn, let, mut, struct, impl, unsafe, Option, Some, None},
  sensitive=true,
  morecomment=[l]{//},
  morestring=[b]{"},
}

\def\System{\textsc{LiteRSan}}
\def\Variant{\textsc{Semi-LiteRSan}}

\newif\ifprintcomments

\newcommand{\kaiming}[1]{
    \ifprintcomments
        \textcolor{red}{{\em\bf Kaiming:} #1}
    \fi}

\printcommentstrue

\def\Comment#1{{\textsl{\color{red}  $\langle\!\langle$#1$\rangle\!\rangle$}} }
\def\JZ#1{\Comment{{\textbf {JZ}}: #1}}

\pgfplotsset{compat=1.9}

\DeclareMathAlphabet{\mathcal}{OMS}{cmsy}{m}{n}

\DeclareGraphicsExtensions{%
    .png,.PNG,%
    .pdf,.PDF,%
    .jpg,.mps,.jpeg,.jbig2,.jb2,.JPG,.JPEG,.JBIG2,.JB2}

\usepackage{xparse}
\newcommand{\bnm}{\begin{newmath}}
\newcommand{\enm}{\end{newmath}}

\newcommand{\bea}{\begin{eqnarray*}}%
\newcommand{\eea}{\end{eqnarray*}}%

\newcommand{\bne}{\begin{newequation}}
\newcommand{\ene}{\end{newequation}}

\newcommand{\bal}{\begin{newalign}}
\newcommand{\eal}{\end{newalign}}

\newenvironment{newalign}{\begin{align}%
\setlength{\abovedisplayskip}{4pt}%
\setlength{\belowdisplayskip}{4pt}%
\setlength{\abovedisplayshortskip}{6pt}%
\setlength{\belowdisplayshortskip}{6pt} }{\end{align}}

\newenvironment{newmath}{\begin{displaymath}%
\setlength{\abovedisplayskip}{4pt}%
\setlength{\belowdisplayskip}{4pt}%
\setlength{\abovedisplayshortskip}{6pt}%
\setlength{\belowdisplayshortskip}{6pt} }{\end{displaymath}}

\newenvironment{newequation}{\begin{equation}%
\setlength{\abovedisplayskip}{4pt}%
\setlength{\belowdisplayskip}{4pt}%
\setlength{\abovedisplayshortskip}{6pt}%
\setlength{\belowdisplayshortskip}{6pt} }{\end{equation}}

\newcounter{ctr}

\newcounter{mytable}
\def\mytable{\begin{centering}\refstepcounter{mytable}}
\def\endmytable{\end{centering}}

\newcounter{myfig}
\def\myfig{\begin{centering}\refstepcounter{myfig}}
\def\endmyfig{\end{centering}}

\newlength{\saveparindent}
\newlength{\saveparskip}
\newcommand{\E}{{\rm I\kern-.3em E}}

\renewcommand{\eqref}[1]{\mbox{Equation~(\ref{#1})}}

\renewcommand{\u}{\ensuremath{u}}

\def \part {part}

\renewcommand{\paragraph}[1]{\vspace*{3pt}\noindent\textbf{#1}\;}

\def \blackslug{\hbox{\hskip 1pt \vrule width 4pt height 8pt
    depth 1.5pt \hskip 1pt}}
\def \qed{\quad\blackslug\lower 8.5pt\null\par}

\newcounter{mynote}[section]

\newcommand\ignore[1]{}

\newcounter{rcnote}[section]

\newcounter{mrnote}[section]

\newcounter{fknote}[section]

\newcounter{anote}[section]

\DeclareMathSymbol{\mlq}{\mathord}{operators}{``}
\DeclareMathSymbol{\mrq}{\mathord}{operators}{`'}

\newcommand{\rhf}[2]{R_{f, \gamma}}

\DeclareDocumentCommand{\edist}{o o}{
  \ensuremath{
    \IfNoValueTF{#1}{{d}}{{\sf d}(#1,#2)}
  }
}

\newcommand{\olrk}[1]{\ifx\nursymbol#1\else\!\!\mskip4.5mu plus 0.5mu\left(\mskip0.5mu plus0.5mu #1\mskip1.5mu plus0.5mu \right)\fi}

\NewDocumentCommand{\indseq}{ O{1} O{r} }{{#1}\ldots {#2}}

\begin{document}
\def\thetitle{{\System}: Lightweight Memory Safety Via Rust-specific Program Analysis and Selective Instrumentation}
\title{\thetitle}

\author{Tianrou Xia}
\affiliation{The Pennsylvania State University}
\email{tzx17@psu.edu}

\author{Kaiming Huang}
\affiliation{The Pennsylvania State University}
\email{kzh529@psu.edu}

\author{Dongyeon Yu}
\affiliation{UNIST}
\email{dy3199@unist.ac.kr}

\author{Yuseok Jeon}
\affiliation{Korea University}
\email{ys_jeon@korea.ac.kr}

\author{Jie Zhou}
\affiliation{The George Washington University}
\email{jie.zhou@gwu.edu}

\author{Dinghao Wu}
\affiliation{The Pennsylvania State University}
\email{dinghao@psu.edu}

\author{Taegyu Kim}
\affiliation{The Pennsylvania State University}
\email{tgkim@psu.edu}

\date{}

\begin{abstract}
Rust is a memory-safe language, and its strong safety guarantees combined with high performance have been attracting widespread adoption in systems programming and security-critical applications. However, Rust permits the use of \emph{unsafe} code, which
bypasses compiler-enforced safety checks and can introduce memory vulnerabilities. 
A widely adopted approach for detecting memory safety bugs in Rust is Address Sanitizer (ASan).
Optimized versions, such as ERASan and RustSan, have been proposed to selectively apply security checks in order to reduce performance overhead.
However, these tools still incur significant performance and memory overhead and fail to detect many classes of memory safety vulnerabilities due to the inherent limitations of ASan.

In this paper, we present \System{}, a novel memory safety sanitizer that
addresses the limitations of prior approaches. By leveraging Rust's unique 
ownership model, \System{} performs Rust-specific static analysis that is aware of pointer lifetimes to identify risky pointers. It then selectively instruments risky pointers to enforce only the 
necessary spatial or temporal memory safety checks.
Consequently, {\System} introduces 
significantly lower runtime overhead (18.84\% versus 152.05\% and 183.50\%) and negligible memory overhead
(0.81\% versus 739.27\% and 861.98\%) compared with existing ASan-based sanitizers
while being capable of detecting memory safety bugs that prior techniques miss.

\end{abstract}

\maketitle


\section{Introduction}
\label{sec:introduction}

Memory-safe programming languages have emerged as a promising approach~\cite{whitehousememsafe} to mitigate prevalent memory safety vulnerabilities, which account for 70\%--80\% of all software vulnerabilities~\cite{nsa-report,microsoft-report,google-report}. 
Among these languages, Rust~\cite{rust:book2022} stands out by enforcing strong compile-time safety guarantees. Its advanced type system detects security issues early and helps confine additional costs to protect safety-critical operations. 
Studies show that, aside from these checks, Rust’s performance can closely match that of C/C++~\cite{RustPerf:ASE22}. 
Consequently, Rust has rapidly gained adoption in security-critical and performance-sensitive domains~\cite{Google:Rust2021,AWS:Rust2022,Servo:2019}.

Despite its robust safety guarantees, Rust's type system is not flawless. 
It can be too restrictive, 
\hl{preventing the expressiveness required for low-level systems programming, 
or it may introduce prohibitive runtime overhead in performance-critical code paths.}
Consequently, Rust permits \emph{unsafe} code, such as raw pointer dereferences or calling external C library
functions~\cite{UnsafeRust:ICSE20,UnsafeRust:OOPSLA20,UnsafeRust:TOSE24},
enabling developers to bypass Rust's memory safety checks.
Nevertheless, the use of unsafe Rust code 
\hl{reintroduces} memory safety vulnerabilities, such as buffer overflows and 
Use-After-Free (UAF) bugs, undermining Rust's foundational memory-safety
benefits~\cite{MemSafetyRust:TOSEM21,UnsafeRust:TOSE24,RustCVE:2018-1000810,RustCVE:2019-16760}.

Various detection and mitigation mechanisms have been proposed to address memory safety
challenges introduced by unsafe Rust. Static analysis tools, such as
Rudra~\cite{Rudra:SOSP21}, MirChecker~\cite{MirChecker:CCS21}, and
SafeDrop~\cite{SafeDrop:TOSEM23}, have successfully identified many real-world
vulnerabilities in Rust programs. However, these tools typically suffer from 
high false positives (e.g., Rudra reports approximately 89\% false
positives~\cite{Rudra:SOSP21}), and have limited capability in detecting diverse
bug types~\cite{ERASan:Oakland24,RustSan:SEC24}.  Memory isolation techniques,
such as XRust~\cite{XRust:ICSE20}, TRust~\cite{TRust:Sec23} and PKRUSafe~\cite{PKRUSafe:EuroSys22},
provide runtime protection by restricting unsafe code's access to memory
objects exclusively used by safe code. Nonetheless, these approaches target 
only subsets of memory objects and primarily focus on spatial memory errors
(e.g., buffer overflows) while neglecting temporal memory errors such as UAF.  
Rust fuzzing
frameworks~\cite{web:afl.rs,web:cargo-fuzz,web:honggfuzz-rs,RPG:ICSE24} have
also emerged to detect memory safety vulnerabilities. However, it is well-known that the probabilistic nature of the fuzzing approach results in challenges of \hl{systematical detection} of memory errors~\cite{FuzzResidualRisk}.
%





Researchers have also developed tools based on Address Sanitizer
(ASan)~\cite{ASan:ATC12}---a compiler-based memory error detector---to reveal
memory safety vulnerabilities in Rust.
Compared to static analysis (limited bug detection capability),
memory isolation (partial protection), and fuzzing (probabilistic by nature), 
ASan-based approaches provide deterministic dynamic validation of
\emph{every} memory access.  Notably,
ERASan~\cite{ERASan:Oakland24} and RustSan~\cite{RustSan:SEC24} have advanced this area by optimizing away
redundant checks for memory accesses already instrumented by
the Rust compiler, thereby significantly reducing
ASan's runtime overhead by 71.4\% and 62.3\%, respectively.

Despite these advances, existing ASan-based tools still do not fully align with
Rust’s native safety guarantees. Although ERASan and RustSan remove certain 
checks already enforced by Rust’s type system, their reliance on traditional C/C++ pointer analyses (i.e., SVF~\cite{SVF:CC16}) leads to significant over-approximation of
unsafe pointers, as such analyses are not integrated with Rust's ownership and borrowing semantics~\cite{rustownership}.  As a result, both tools introduce superfluous checks for 
memory accesses that are already guaranteed to be safe, imposing 
unnecessary runtime overhead.
In addition, the static analysis time of ERASan and RustSan is prohibitively high, 
increasing compilation time by 1,635.35\% and 1,193.31\% per our measurements, due to their reliance on SVF, 
which is particularly expensive for large programs~\cite{PKRUSafe:EuroSys22}.
Furthermore, ASan suffers from inherent limitations in bug detection.
Its red zones can be 
bypassed by overflows that exceed the boundaries, and its shadow memory mechanism may fail to detect UAF bugs when freed memory is reallocated post-quarantine, which makes dangling pointers to the original object undetectable.
These gaps cause ASan-based tools to provide incomplete bug coverage despite significant overhead.


To address these limitations, our goal is to design a memory error detection mechanism tailored to Rust’s inherent safety guarantees while addressing the loopholes introduced by unsafe code and the weaknesses of existing detection frameworks. Specifically, we strive to (1) identify pointers that truly pose spatial or temporal risks by incorporating a Rust-specific static analysis, eliminating extraneous checks on pointers that are either statically-proven safe or protected with compiler-inserted checks, (2) maintain complete coverage and precision in detecting {\bf\em all} classes of memory errors, including spatial and temporal errors without using heavyweight ASan-based approaches, and (3) minimize overhead by integrating Rust’s ownership and borrowing rules into both static analysis and enforcing selective instrumentation for lightweight runtime checks.

Achieving these three objectives requires addressing three major challenges: (1) Rust’s allowance of raw pointers within otherwise safe code complicates standard pointer analysis, as many may-alias inferences valid in the C/C++ context break under Rust’s stricter ownership model, (2) bridging static checks and runtime validation demands a lightweight metadata design that captures Rust memory safety model, and (3) avoiding expensive and coarse-grained ASan-based runtime checks. To address these challenges, we developed \System{} (\textbf{\textsc{Lite}}-\textbf{\textsc{R}}ust-\textbf{\textsc{San}}itizer), which deploys a Rust-specific static analysis to pinpoint \emph{truly risky} pointers and selectively instrument them with minimal metadata to detect both spatial and temporal memory errors at runtime. This synergy of compile-time insights and targeted runtime checks enables comprehensive and accurate memory error detection with minimal overhead across 28 widely used Rust benchmarks: only 18.84\% runtime, 0.81\% memory and 97.21\% compile-time overhead. In contrast, ERASan incurs 152.05\% runtime, 739.27\% memory, and 1,635.35\% compile-time overhead, while RustSan incurs 183.50\%, 861.98\%, and 1,193.31\%.

In summary, we make the following contributions:
\begin{itemize}[leftmargin=*, topsep=1pt, itemsep=1pt] 
\item \textbf{Rust-specific Taint Analysis:} We introduce a Rust-specific static analysis scheme that identifies risky pointers by integrating Rust’s ownership and borrowing semantics rather than defaulting to generic pointer analysis.

\item \textbf{Lightweight Metadata Inference and Runtime Checks:} We design a compact metadata mechanism for runtime validation of spatial and temporal safety, removing the heavyweight components (e.g., red zones and shadow memory) of classic sanitizers.

\item \textbf{Comprehensive and Efficient Bug Detection:} Our approach, \System{}, systematically detects spatial and temporal memory errors in Rust. Compared to prior Rust sanitizers, \System{} offers complete coverage and higher accuracy in detecting bugs while minimizing compile-time, runtime, and memory overhead compared with existing ASan-based tools.

\end{itemize}

\section{Background}
\label{sec:background}

In this section, we explain the background on Rust’s safety guarantees and root causes of memory safety violations (\S\ref{sec:bg:rust}). We also briefly discuss pointer analyses and AddressSanitizer (ASan)~\cite{ASan:ATC12}, which are commonly used by existing Rust memory safety tools, along with their limitations (\S\ref{sec:bg:ptr} and \S\ref{sec:bg:asan}). Finally, we present a
motivating example to illustrate \hl{the redundant checks applied by prior ASan-based sanitizers} (\S\ref{sec:motivation}).
%

\subsection{Rust's Memory Safety Guarantee}
\label{sec:bg:rust}

\paragraph{Spatial memory safety.} 
Rust prevents out-of-bounds memory accesses by disallowing explicit pointer arithmetic on references and by internally maintaining spatial metadata (e.g., capacity and length) for containers, such as vector~\cite{rust:book2022}. At compile time, the compiler either verifies the safety of a memory dereference or inserts runtime checks to detect any out-of-bounds access. However, Rust also permits \emph{unsafe} code regions in which developers can manipulate raw pointers directly. These unsafe constructs bypass the compiler’s spatial checks and can lead to memory safety violations, making them the root cause of out-of-bounds errors~\cite{UnsafeRust:OOPSLA20,UnsafeRust:TOSE24,ERASan:Oakland24}.

\paragraph{Temporal memory safety.} 
Through its ownership and borrowing model, Rust ensures that each memory object has a single owner and that all borrowed references remain valid only as long as that owner is in scope~\cite{rust:book2022}. This prevents use-after-free and double free by enforcing deallocation once the owner goes out of scope. Nevertheless, \emph{unsafe} code regions allow the creation and handling of raw pointers in ways that can violate the ownership rules, enabling temporal errors if these pointers outlive their underlying objects. As with spatial safety, these unsafe constructs are the primary source of temporal memory safety violations.

\subsection{Pointer Analyses in Rust}
\label{sec:bg:ptr}

Prior work on identifying unsafe pointers (i.e., those that may lead to memory errors) in Rust typically relies on classical alias analysis~\cite{IsPASolved:PASTE01}.
In C/C++ contexts, such analysis often produces over-approximation: when it cannot disprove aliasing between two pointers, the analysis labels them as {\em may-alias}~\cite{intermayalias,aliasnphard}. This approach does not account for Rust's ownership and borrowing rules, where each object has a unique owner and references are strictly managed. Consequently, it leads to unnecessary and even higher false positives in Rust context compared with C/C++, as many pointers flagged are actually safely managed by the Rust compiler.

A prominent example is SVF~\cite{SVF:CC16}, a state-of-the-art pointer alias analysis tool, used by both ERASan~\cite{ERASan:Oakland24} and RustSan~\cite{RustSan:SEC24}. While SVF supports sophisticated inter-procedural and context-sensitive analyses, it suffers from two key drawbacks when applied to Rust. First, SVF’s alias analysis significantly over-approximates due to its inability to leverage Rust’s strict ownership semantics, resulting in the imprecise identification of unsafe pointers. Second, SVF incurs substantial analysis time, particularly on medium to large code bases~\cite{hybridglobalalias,PKRUSafe:EuroSys22}. While powerful in theory, SVF introduces considerable computational overhead and scalability issues, making it impractical for large-scale analysis pipelines.

\subsection{Address Sanitizer and Its Limitations}
\label{sec:bg:asan}

Address Sanitizer (ASan)~\cite{ASan:ATC12} is a widely adopted tool for detecting memory safety violations at runtime, including both spatial errors and temporal errors.
Its practicality and effectiveness have led to broad integration across major compilers and use in projects written in C, C++, and Rust. 
ASan instruments each memory access instruction with runtime checks to validate its legitimacy.
%

ASan detects memory errors using three core mechanisms: \emph{red zones}, \emph{shadow memory}, and \emph{quarantine}. 
However, ASan’s red zones are limited in size, as large overflows that bypass ASan's red zones evade detection.
Meanwhile, once a memory region is freed and subsequently reallocated, the shadow memory is updated for the new allocation, erasing the evidence of original dangling pointers.
Even with the quarantine mechanism that temporarily delays the reuse of
freed memory \hl{regions} by placing them in a quarantine pool, this protection is
short-lived.
Thus, ASan may miss temporal violations if a quanrantined region is reallocated
while dangling pointers to the region are still in scope.

In addition to its incomplete coverage, ASan introduces substantial runtime and memory overhead. 
Typical performance slowdown ranges from 2--3$\times$, and the red zone and shadow memory can cause the overall memory overhead to grow by several times.
These limitations highlight the need for a more precise and lightweight memory safety mechanism. 
Ideally, such a mechanism would avoid red zones and shadow memory while preserving strong detection capabilities for both spatial and temporal safety violations.


\subsection{Motivating Example}
\label{sec:motivation}

While Rust enforces memory safety for most memory accesses
(\S\ref{sec:bg:rust}), severe errors (e.g., buffer overflows and 
UAF) can still occur when unsafe code is used.  
Listing~\ref{lst:uaf_cache} shows an example of a common scenario in web
applications (e.g., Servo~\cite{Servo:2019}).

\begin{figure}[h]
\centering
\begin{lstlisting}[abovecaptionskip=1pt,belowcaptionskip=0pt,numbersep=5pt,xleftmargin=1.3em,caption={Use-after-free by caching a raw pointer after ownership transfer.},label={lst:uaf_cache}]
struct Cache {
    ptr: Option<*mut u8>,
}

impl Cache {
    fn save(&mut self, ptr: *mut u8) {
        self.ptr = Some(ptr);}

    fn load(&self) -> Box<String> {
        unsafe {
            Box::from_raw(self.ptr.unwrap())}}
}

fn main() {
    let mut cache = Cache { ptr: None };
    let token = Box::from("session-token");
    println!("Session token: {}", token);
    {
        let local_token = token;
        cache.save(local_token.as_ptr() as *mut u8);
        // local_token goes out of scope here.
    }
    let stale_token = cache.load(); // Dangling pointer
    println!("Stale session token: {}", stale_token); // UAF
}
\end{lstlisting}
\vspace{-5pt}
\end{figure}

In Listing~\ref{lst:uaf_cache}, the string {\tt ``session-token''} is a
heap object allocated at line~16. A smart pointer, \texttt{token}, points to 
and owns this object. At line~19, ownership is transferred: a new smart 
pointer, \texttt{local\_token}, takes the ownership of the heap object. 
Lines~20 and~7 define a raw pointer, \texttt{self.ptr}, derived
from \texttt{local\_token}. This raw pointer does not take the ownership of
the string, so the owner remains \texttt{local\_token}.
\texttt{local\_token} goes out of scope at the end of line~21, causing 
the object it owns to be deallocated.
The raw pointer \texttt{self.ptr} then becomes dangling.
At line~23, a new smart pointer, \texttt{stale\_token}, is created from
the dangling raw pointer, and it also becomes dangling.
Then, all subsequent dereferences of the two dangling pointers
are UAF (e.g., the one at line~24).

ASan instruments all memory accesses, which can be redundant, 
incurring high performance and memory overhead without guaranteeing
comprehensive memory safety.\footnote{As
evaluated in MSET~\cite{sanitizersok25oakland}, ASan failed to detect around
50\% of C/C++ memory errors in their constructed benchmark. We believe the
rationale would be similar for Rust, as ASan is not aware of Rust's
memory safety model.} 
For the example in Listing~\ref{lst:uaf_cache}, no spatial memory safety
check is necessary.
For temporal memory safety, the dereference of {\tt token} (line~17) does not
require safety instrumentation, as Rust’s ownership model ensures its
validity. 

To address this deficiency, prior work---namely, ERASan~\cite
{ERASan:Oakland24} and RustSan~\cite{RustSan:SEC24}---improves ASan's 
performance by selectively instrumenting only raw pointers, or pointers in unsafe code, and their aliases.
However, for the example discussed here, conventional alias analysis would
identify all pointers in Listing~\ref{lst:uaf_cache} as aliases to the raw pointer \texttt{self.ptr} in unsafe code, resulting in redundant checks inserted to the dereference site of a safe pointer (e.g., line~17).
This redundancy stems from insufficient consideration of Rust’s memory 
safety guarantees, leading to over-approximating and instrumenting safe pointer 
dereferences.

\subsection{Ideal Memory Error Detection for Rust}
\label{sec:bg:ideal}
Ideally, a Rust memory safety sanitizer should:
(1) leverage Rust’s memory safety model to precisely differentiate safe pointers (e.g., \texttt{token}) from 
unsafe ones (e.g., \texttt{self.ptr});
(2) selectively instrument only unsafe pointers, avoiding redundant checks on Rust-guaranteed safe pointers; and
(3) provide comprehensive, accurate, and efficient detection of all memory error classes.
Such an approach narrows safety checks to only unsafe 
operations, incurring minimal overhead while ensuring comprehensive detection coverage.

\vspace{10pt}
\section{Threat Model and Challenges}
\label{sec:threatchallenge}

In this section, we introduce the threat model and the challenges to be addressed through the design and implementation of \System{}.

\subsection{Threat Model}
\label{sec:threat_model}

\begin{figure}[t]
  \centering
  \includegraphics[width=0.75\linewidth]{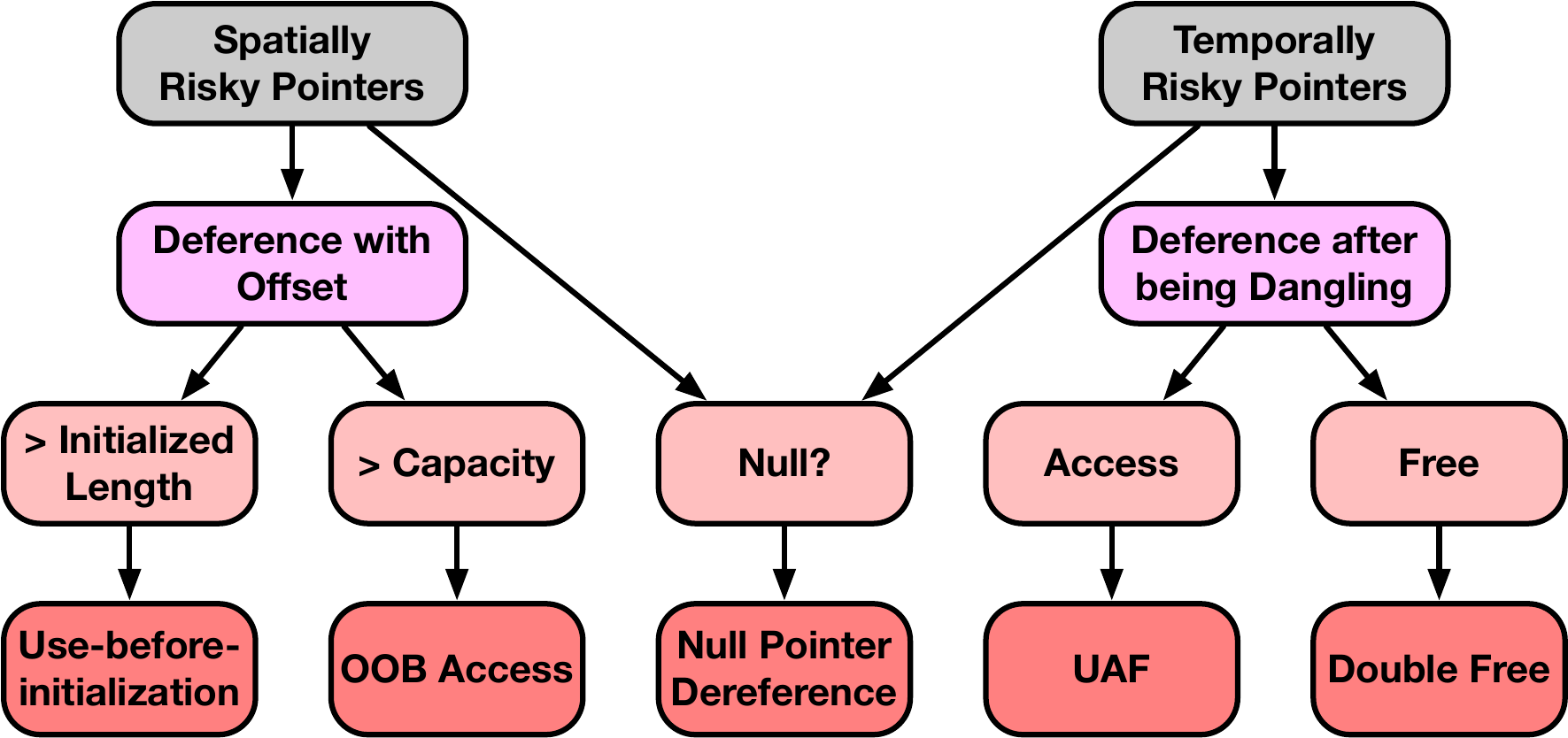}
  \caption{Memory safety bug patterns in Rust programs.}
  \vspace{10pt}
  \label{fig:bug_pattern}
  \vspace{-5pt}
\end{figure}

We assume that memory errors, including both spatial (e.g., 
out-of-bound read/write) and temporal (e.g., UAF and double-free) errors, are possible in Rust programs.
Our goal is to detect all such memory errors.
While directly-linked C/C++ libraries may also contain memory errors,
we focus on Rust source code and neither analyze nor harden such external libraries.  
We also assume that no extra memory safety defenses are deployed beyond Rust’s built-in safety support.
Memory leaks are out of scope, as they are generally not classified as memory safety violations~\cite{6547101, ASan:ATC12, softbound,CETS:ISMM10}.
Figure~\ref{fig:bug_pattern} shows the complete set of the bug patterns that \System{} covers, as ASan-based tools~\cite{ERASan:Oakland24,RustSan:SEC24} do in Rust programs.

%

Non-memory-safety errors, such as concurrency bugs and logic errors, are
outside the protection scope of \System{}.
In addition, \System{} is not designed to detect type conversion bugs.
Notably, Rust's {\tt std::mem::transmute()}~\cite{api:transmute}
allows converting the type of an object to any other type.
{\System} does not address errors caused by misusing this dangerous API.
However, \System{} can detect type confusion bugs that arise from temporal errors, such as UAF.
As mentioned above, {\System} does not target external C/C++ libraries.
Therefore, cross-language attacks~\cite{crosslanguageattack} that 
propagate exploitation from components written in unsafe
languages (e.g., C/C++) are out of scope and can be addressed by existing works~\cite{Galeed:ACSAC21}.

\subsection{Challenges}
\label{subsec:challenge}

As discussed in \S\ref{sec:introduction} and \S\ref{sec:background},
existing memory error detection approaches are both incomplete and 
inefficient. Static analyzers~\cite{Rudra:SOSP21,MirChecker:CCS21} often produce
a high number of false positives, while ASan-based techniques~\cite{ERASan:Oakland24, RustSan:SEC24} incur significant performance overhead and still fail to detect many bugs.
We observed that the shortcomings of ASan-based tools largely
stem from analyzing Rust in LLVM IR~\cite{LLVM:CGO04}---a 
language-independent, low-level compiler intermediate
representation, using generic pointer analysis~\cite{SVF:CC16} without accounting for Rust's unique memory safety guarantees.

To propose our approach, we first introduce the key concept of \emph{risky pointer}, which will be used throughout the rest of this paper.

{\setlength{\parindent}{0pt}
\begin{definition}
A \textbf{\emph{risky pointer}} is a pointer whose dereferences may violate memory safety. Such a pointer is \emph{spatially risky} if it bypasses Rust’s spatial enforcements, or \emph{temporally risky} if it may outlive its referenced object.
\label{def:riskyptr}
\end{definition}
}

Detailed explanations of spatially and temporally risky pointers are presented in \S\ref{sec:riskptr_def}.
Note that (1) a pointer may be both spatially and temporally risky; (2) A raw pointer becomes risky \emph{only when it is exposed in unsafe code} (Definition~\ref{def:exposedrawptr});
and (3) Rust's native smart pointers may also be risky. For example,
constructing multiple smart pointers from a raw pointer may violate Rust’s ownership rules, rendering these smart pointers risky and potentially causing UAF bugs that elude compiler checks.

{\setlength{\parindent}{0pt}
\begin{definition}
An \textbf{\emph{exposed raw pointer}} is a raw pointer \emph{directly used in unsafe code}, bypassing Rust’s safety guarantees.
\label{def:exposedrawptr}
\end{definition}
}

We identify three key challenges in building an efficient and comprehensive memory \hl{safety} sanitizer tailored to Rust.

\begin{itemize}[leftmargin=*, topsep=2pt, itemsep=3pt] 
\item \textbf{C1: Leveraging Rust’s unique type system to precisely
identify risky pointers.}
Program analysis for Rust in prior
work~\cite{ERASan:Oakland24,RustSan:SEC24,TRust:Sec23,XRust:ICSE20} does
not utilize Rust’s ownership and borrowing semantics, significantly
over-approximating risky pointers.
A refined approach should integrate Rust’s intrinsic memory
safety model to more precisely identify risky pointers.

\item \textbf{C2: Managing lightweight safety metadata for runtime checks.}
Relying on a coarse-grained protection scheme like ASan’s shadow memory
and red zones~\cite{ASan:ATC12} is expensive and imprecise.
Tailoring compact \hl{yet fine-grained} metadata that incorporates Rust’s
memory safety guarantees enables more efficient and accurate runtime
error detection. Additionally, because raw pointers lack spatial metadata
(i.e., bounds information), {\System} must infer and maintain their metadata 
\hl{to enable} runtime validation.

\item \textbf{C3: Minimizing overhead while ensuring coverage.}
As Rust’s memory safety model already protects a substantial amount of memory
accesses, additional checks are only needed for those involving risky pointers.
The challenge is to minimize cost while maintaining accuracy and
comprehensiveness by (1) selectively instrumenting only the truly risky pointers
based on their specific risk types and (2) enforcing an efficient runtime check
mechanism rather than incomplete and inefficient ASan-style checks.

 
\end{itemize}

By addressing these challenges, {\System} complements Rust’s inherent memory
safety guarantees with precise instrumentation to achieve comprehensive and
low-overhead runtime safety checks, closing the gap left by prior
work~\cite{ERASan:Oakland24,RustSan:SEC24}.

\vspace{5pt}
\section{Overview}
\label{sec:overview}

\begin{figure*}[t]
  \centering
  \includegraphics[width=0.9\linewidth]{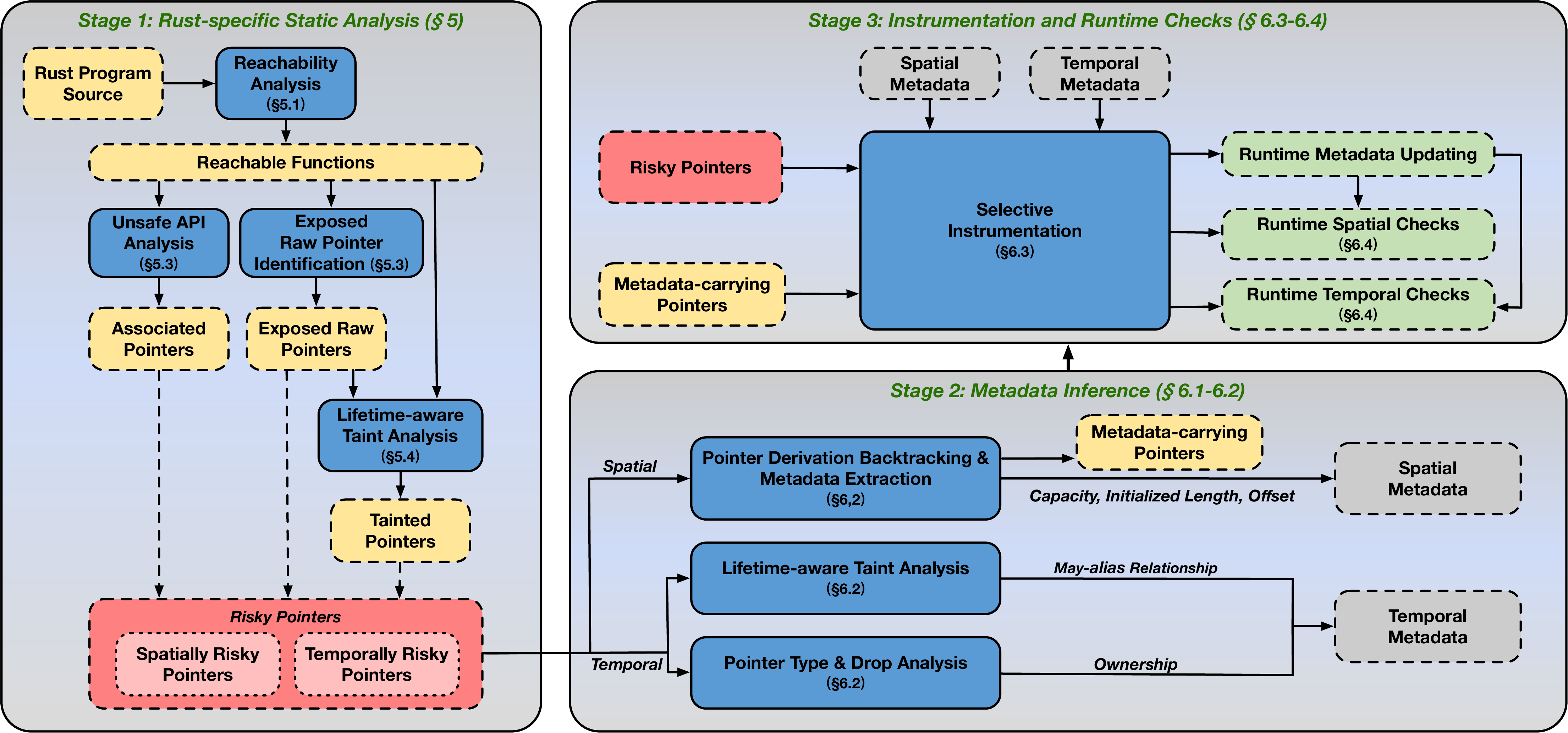}
    \caption{\textbf{\System{} overview.} \System{} consists of three stages. Each addresses one of the primary challenges in enabling efficient and comprehensive sanitizer checks. The output of each stage serves as the input to the next.}
    \vspace{5pt}
  \label{fig:workflow}
\end{figure*}

To address the three key challenges described in \S\ref{subsec:challenge},
we propose a Rust-specific static analysis
to identify risky pointers. Coupling it with a metadata-based runtime checking mechanism,
we develop our prototype system, \System{}.
Figure~\ref{fig:workflow} illustrates the main components and the
overall workflow of \System{}.

\textbf{Stage 1} conducts Rust-specific static analysis to addresses \textbf{C1}. \System{} first pre-processes the target Rust program using reachability analysis to narrow the analysis scope to potentially reachable functions (\S\ref{sec:ana_scope}). Within this scope, \System{} identifies both \emph{Spatially Risky Pointers} (\S\ref{sec:spa_risk}) and \emph{Temporally Risky Pointers} (\S\ref{sec:temp_risk}). 
Since the misuse of exposed raw pointers is the primary cause of memory safety violations in Rust, \System{} begins by identifying them. It annotates the instructions involving raw pointers during Mid-level IR (MIR)~\cite{rust:mir} to LLVM IR code generation, and analyzes the definitions and uses of these pointers in annotated instructions to identify exposed raw pointers, 
which are classified as both spatially and temporally risky because they
are exempt from Rust’s compile-time safety enforcement. To identify additional
temporally risky pointers, \System{} performs \emph{lifetime-aware} taint
analysis starting from exposed raw pointers. 
This is necessary because exposed raw pointers can propagate temporal risks to other pointers referencing the same memory object.
In contrast, spatial risks do not propagate if raw pointer arithmetic and dereference is instrumented with bounds checking. 
Additionally, \System{} identifies risky pointers used in unsafe APIs that may cause memory safety violations.


\textbf{Stage 2} constructs lightweight spatial and temporal metadata to address
\textbf{C2} (\S\ref{sec:metadata_def} \S\ref{sec:metadata_inference}), enabling
efficient runtime validation. For spatially risky pointers, \System{} maintains
three pieces of metadata: capacity, initialized length and offset. When spatial
metadata is unavailable at a pointer’s definition site (e.g., an exposed raw pointer
derived from other pointers), \System{} backtracks pointer derivations (Definition~\ref{def:derivation}) to
extract metadata from the object’s allocation site. This process also identifies
\emph{metadata-carrying pointers}, which are responsible for transmitting
spatial metadata at runtime.

For temporally risky pointers, the risk arises from shared access to the same
memory object. Once the object is deallocated by its owner or via an unsafe API,
all referencing pointers that remain in scope become dangling. To address this,
\System{} maintains may-alias relationships and ownership information as
temporal metadata.  May-aliases are established during Stage~1 via taint
analysis, as pointers tainted by the same exposed raw pointer reference the same
object. Among these, \System{} analyzes pointer types and \texttt{Drop}
implementations~\cite{Drop:Rust} to identify owners.

%

\textbf{Stage 3} addresses \textbf{C3} by selectively instrumenting identified
risky pointers and metadata-carrying pointers (\S\ref{sec:instrumentation}),
thereby minimizing runtime overhead while preserving the comprehensive coverage
of runtime checks. Leveraging spatial and temporal metadata
collected in Stage 2 and updating it during execution, \System{} performs
accurate and comprehensive detection of both spatial and temporal memory errors
(\S\ref{sec:runtime_check}). The complete set of memory safety bugs detectable by
\System{} is summarized in Figure~\ref{fig:bug_pattern}.

\section{Rust-Specific Static Analysis}
\label{sec:design}

In this section, we present \System{}'s Rust-specific static analysis,
which addresses the challenge of identifying risky pointers (C1) discussed in
\S\ref{subsec:challenge}.  We begin by defining the scope of the analysis in
\S\ref{sec:ana_scope} and introducing the definition of risky pointers in
\S\ref{sec:riskptr_def}.  We then describe our approach to identifying
spatially and temporally risky pointers in \S\ref{sec:spa_risk} and
\S\ref{sec:temp_risk}, respectively, and conclude this section by discussing
soundness and precision in \S\ref{sec:soundness}.

\subsection{Static Analysis Scope Restriction}
\label{sec:ana_scope}

\System{} restricts its static analysis to \emph{reachable functions},
motivated by the structure of Rust programs, which often include deeply nested
library code, much of which is dead code (i.e., unreachable from the program
entry point\footnote{The entry point is typically the \texttt{main} function.
For library crates compiled with built-in benchmarks, benchmarking functions compiled and included in LLVM IR are treated as entry points as well.}). To exclude such dead code from
{\System}'s analysis, \System{} performs \emph{reachability analysis} to
conservatively identify and analyze only potentially reachable functions \hl{which will be executed} at
runtime.

Specifically, starting from the program’s entry point, \System{} identifies
and enqueues both directly called functions and address-taken functions (i.e.,
potential indirect call targets~\cite{typedive,deeptype}) for analysis. For each
function in the queue, \System{} recursively discovers and further enqueues the
function's callees and address-taken functions, thereby restricting its analysis
scope to functions potentially reachable during execution.  By limiting analysis
to reachable functions, \System{} focuses on identifying risky pointers within
this scope that may lead to memory safety violations.

\subsection{Risky Pointer Definition}
\label{sec:riskptr_def}

Within {\System}'s restricted analysis scope, most pointers are safe thanks to
Rust’s native safety guarantees for safe code, as discussed in
\S\ref{sec:bg:rust}.  However, a subset of pointers remains unprotected and may
still violate memory safety. We refer to these as \emph{risky pointers}
(Definition~\ref{def:riskyptr}), and \System{} focuses its safety checks
exclusively on them.  For fine-grained analysis and instrumentation, we further
classify risky pointers into \emph{spatially risky} and \emph{temporally
risky} categories, corresponding to potential violations of spatial and temporal
memory safety, respectively.

\textbf{Spatially risky pointers} include (1) exposed raw pointers and (2) smart
pointers used in certain unsafe APIs. Unlike encapsulated raw pointers within smart pointers,  which enforce Rust's safety guarantees, exposed 
raw pointers are spatially unsafe because arbitrary pointer arithmetic is permitted on them, which may result in invalid
pointers whose dereferences are \hl{out-of-bound and} not checked. Moreover, Rust’s standard
libraries (e.g. \texttt{std}) provide unsafe APIs that may subvert bounds
checking if misused~\cite{MetaSafe:Sec24,rustunsafeapi}. When a pointer is used
in conjunction with such unsafe APIs, it is considered spatially risky.

\textbf{Temporally risky pointers} include (1) exposed raw pointers and (2) any
\emph{valid} (i.e., in-scope according to Rust's scoping
rules~\cite{rust:scope}) pointers that reference the same memory object as \hl{an exposed} raw
pointer. Exposed raw pointers are temporally unsafe because they are exempt from Rust’s
ownership rules; once the referenced object is deallocated, such raw pointers
become dangling. Furthermore, as illustrated in \S\ref{sec:motivation}, if a
smart pointer is constructed from an exposed raw pointer and takes ownership of an object
that already has an owner, multiple owners will coexist. Deallocating the
object through one owner leaves the others dangling.  Invalid pointers whose lifetimes have ended (e.g.,
\texttt{token} in Listing~\ref{lst:uaf_cache}) are excluded from temporally
risky pointers, since any use of them is prevented by Rust compiler.

\subsection{Spatially Risky Pointer Identification}
\label{sec:spa_risk}

\paragraph{Exposed raw pointers.}
The misuse of exposed raw pointers is a primary cause of memory safety bugs in Rust programs~\cite{ERASan:Oakland24}. To identify them, \System{} first tracks all raw pointers via LLVM IR metadata annotation during the MIR-to-LLVM IR lowering phase, followed by a fine-grained filtering to determine which are exposed raw pointers, as outlined in \S\ref{sec:overview}. 

\sloppy
Based on ERASan~\cite{ERASan:Oakland24}'s approach, \System{} attaches custom
LLVM metadata~\cite{LLVMLangRef} to IR instructions by modifying the
\texttt{codegen-ssa} and \texttt{codegen-llvm} components of the \texttt{rustc}
compiler.
To determine the locations of annotations, \System{} performs a type-matching
analysis during the MIR-to-LLVM IR lowering phase.  Specifically, it analyzes
the types of program variables and expressions in the MIR (i.e., Rust’s
mid-level representation) to identify those involving raw pointers (e.g.,
\texttt{*const T}). If a value is of raw pointer type, the
corresponding LLVM IR instruction is tagged with \texttt{!rawptr}. Additionally,
instructions originating from unsafe code are marked with
\texttt{!unsafe}.  This analysis allows \System{} to propagate type information
from Rust's MIR and identify the instructions relevant to raw pointers in the
resulting LLVM IR.

After annotating the LLVM IR, \System{} analyzes instructions tagged with \texttt{!rawptr} to filter out encapsulated raw pointers and identify only exposed raw pointers. 
These annotated instructions either define or use raw pointers.
For each definition, \System{} checks whether this raw pointer is used within unsafe code by examining the presence of the \texttt{!unsafe} metadata. If so, it is identified as an exposed raw pointer.
For each use, \System{} similarly checks whether it occurs in unsafe code and, if so, traces back to the corresponding definition site to identify the exposed raw pointer. 
In short, only raw pointers that are directly used within unsafe code are identified for subsequent lifetime-aware taint analysis. 
This filtering is crucial because encapsulated raw pointers within safe abstractions
(e.g., smart pointer creation) are never directly dereferenced and therefore
do not pose risks. Through this analysis, \System{} accurately identifies only
exposed raw pointers that may cause
memory safety bugs, which are considered as
\emph{risky pointers}---both spatially risky and temporally risky (see
\S\ref{sec:riskptr_def}).

\paragraph{Unsafe APIs.} 
As Rust's standard libraries (e.g., {\tt std}) provide unsafe APIs that may
cause spatial safety violations~\cite{MetaSafe:Sec24,rustunsafeapi}, \System{}
analyzes these APIs to extend its protection. In general, an API may be unsafe 
because (1) it directly uses exposed raw pointers or (2) it subverts Rust's bounds 
checks when misused. \System{} can handle type~(1) APIs by identifying
underlying exposed raw pointers using the method discussed above and marking them as
spatially risky. 

Type~(2) APIs are more challenging to address. A notable example is
\texttt{vec::set\_len()}~\cite{api:set_len}, which can alter a vector’s length
to an arbitrary value, potentially resulting in out-of-bounds accesses that
bypass the compiler's spatial safety checks.  Automatically and comprehensively
identifying such APIs would requires analyzing and \emph{understanding} all
library code, which is an undecidable problem~\cite{RiceTheorem}.  Therefore, we
manually examined Rust’s standard libraries and identified nine APIs that may
circumvent bounds checks (Appendix~\ref{sec:appen:api}).  \System{} marks the
pointers involved in these APIs as spatially risky, updates their metadata, and
inserts runtime checks accordingly.  For example, to detect out-of-bounds
accesses potentially caused by \texttt{vec::set\_len()}, \System{} retrieves the
capacity of the vector at its definition site and inserts a spatial check at the
API’s call site to verify whether the new length (i.e., the argument to
\texttt{vec::set\_len()}) exceeds the legal capacity.

\subsection{Temporally Risky Pointer Identification}
\label{sec:temp_risk}


To detect temporal memory safety violations, \System{} must go beyond merely
identifying exposed raw pointers (as discussed in \S\ref{sec:spa_risk}). 
It must also detect any valid pointer that might reference the same memory
object as an exposed raw pointer.
This may sound similar to finding all may-alias pointers to this memory object;
however, the key difference is that any may-alias smart pointers that have
been \emph{moved} or gone out of scope should be excluded
(e.g., {\tt token} after line~19 in Listing~\ref{lst:uaf_cache}).

One approach to identifying temporally risky pointers is to perform alias
analysis on each exposed raw pointer, find the complete set of pointers 
that refer to the same memory object, and then filter out 
aliased pointers that are invalid.
However, existing alias
analysis frameworks for LLVM (e.g. SVF~\cite{SVF:CC16}) conservatively mark two pointers as may-alias whenever they cannot prove that the pointers are not aliases, which leads to over-approximated alias sets even before accounting for the additional over-approximation introduced by ignoring Rust’s safety guarantees.
Furthermore, whole-program alias analysis is generally highly expensive for 
large programs~\cite{PKRUSafe:EuroSys22}.
Therefore, we develop a new {\em Rust-specific, inter-procedural, 
flow-sensitive, lifetime-aware taint analysis} to {\bf directly} identify 
temporally risky pointers without relying on \hl{traditional} alias analysis.

\paragraph{Illustrative example.}
We reuse the example in Listing~\ref{lst:uaf_cache} to briefly illustrate the 
workflow of {\System}'s taint analysis.
In this example, an exposed raw pointer, \texttt{self.ptr} is defined at line 20
through an existing smart pointer (\texttt{local\_token}),
which owns a heap object. Here, the exposed raw pointer \texttt{self.ptr} serves as
a \emph{taint source}. {\System}'s taint analysis performs two key operations
to identify temporally risky pointers to the object pointed by {\tt self.ptr}: 

\begin{itemize}[leftmargin=*, topsep=1pt, itemsep=1pt] 
\item \textbf{Backward propagation}
traces the ownership transfer preceding the definition of the taint source. This includes identifying the smart pointer (\texttt{local\_token})
whose ownership is transferred from \texttt{token}. Since \texttt{token}
goes out of scope before \texttt{self.ptr} is defined, the analysis terminates
backward propagation at \texttt{local\_token}'s definition site (line 19) without tainting \texttt{token}.

\item \textbf{Forward propagation} tracks pointers derived from the taint source and taints them.
In this example, a new pointer
\texttt{stale\_token} is derived from \texttt{self.ptr} (line 23),
making it be tainted and identified as temporally risky.

\end{itemize} 

In short, to capture all temporal safety violations, the taint
analysis must consider both the preceding ownership history of any object
referenced by an exposed raw pointer (i.e., backward) and all subsequent 
pointers derived from that raw pointer (i.e., forward).

\paragraph{Exposed raw pointer classification.} To distinguish the exposed raw pointers (taint sources) that
require different taint analysis propagation directions, we classify them as follow.

\begin{itemize}[leftmargin=*, topsep=1pt, itemsep=1pt] 

\item \textbf{Type 1 (T1) raw pointers:} exposed raw pointers created by
referencing an object already owned by an existing smart pointer (e.g., via
\texttt{Vec::as\_ptr()}~\cite{api:as_ptr}). As shown in Listing~\ref{lst:uaf_cache},
\texttt{self.ptr} is derived from a valid smart pointer \texttt{local\_token}.
The creation of such raw pointers implies existing ownership.
Consequently, T1 raw pointers require both backward taint analysis (to trace the
ownership history of the referenced object) and forward taint analysis (to track
subsequent pointer derivations). 

\item \textbf{Type 2 (T2) raw pointers:} exposed raw pointers created to
reference a newly allocated memory object. Since there is no
preexisting ownership chain to consider, T2 raw pointers require only forward
taint analysis.
\end{itemize} 

After locating these definitions, \System{} performs an inter-procedural taint 
analysis starting from the definition site of each exposed raw pointer (i.e., taint source) and propagating on only forward or both directions according to the class of raw pointers. 

\paragraph{Lifetime-aware taint analysis.} 
\System{} performs a combination of backward and forward taint analysis, \hl{both of which track pointer derivation instructions (Definition~\ref{def:derivation})}, augmented with lifetime-aware propagation that respects Rust’s ownership rules, to identify temporally risky pointers.

{\setlength{\parindent}{0pt}
\begin{definition}
\hl{A \textbf{\emph{pointer derivation instruction}} is any operation that produces a new pointer value from an existing one, by one of the following:
\begin{itemize}
    \item Assignment (direct copy of a pointer),
    \item Computation (arithmetic or type conversion),
    \item Memory propagation (store/load through objects), or
    \item Inter-procedural transfer (via function calls or returns). 
\end{itemize}}
\label{def:derivation}
\end{definition}
}

\begin{itemize}[leftmargin=*, topsep=1pt, itemsep=1pt] 

\item \textbf{Backward taint analysis:} Starting from each taint source (i.e.,
the definition site of a T1 pointer), the analysis traces backward along the
derivation chain to the pointers from which T1 is derived. These pointers share
the same referenced object with T1, and their definition sites are marked as
taint sinks. This propagation halts if ownership is transferred, as any further
use of the original owners is disallowed by the Rust compiler, ensuring that the original owners are free from temporal memory safety violations. In particular, if a
pointer is invalidated before the exposed raw pointer is defined---through function
return for stack pointers or through explicit \texttt{drop} for heap
pointers---it is considered safe and excluded from tainted pointers, as it can
no longer contribute to temporal safety violations. 

\item \textbf{Forward taint analysis:} Starting from each taint source (i.e.,
the definition site of a T1 or T2 pointer), the analysis propagates taint to
all pointers derived from it. Any pointer derived from a tainted
pointer is likewise marked as tainted. This forward propagation continues until
no further \hl{pointer} derivations exist.
\end{itemize}

\paragraph{Inter-procedural analysis.}
Taint propagates inter-procedurally through function calls and returns.
For direct calls, forward propagation flows from actual arguments at the call
site to the corresponding formal parameters of the callee, and from the callee’s
return value to the variable receiving it in the caller. Backward propagation
flows in the reverse direction and terminates at pointers that have been
invalidated, such as those that are out of scope or have transferred ownership.

For indirect calls, \System{} conservatively resolves potential call targets
using a type-based analysis~\cite{tice.etal+14, mcfi}. This approach matches
function signatures (i.e., function prototypes) at indirect call sites with
those of address-taken functions. Although more advanced multi-layer type
analysis techniques~\cite{typedive, deeptype} can improve precision, they
introduce additional static analysis overhead, while the precision gain in
Rust is limited. This is because Rust programs typically rely less on dynamic
dispatch~\cite{rust:dyn_dispatch:1, rust:dyn_dispatch:2, rust:dyn_dispatch:3}
than programs in other languages, and the multi-layered structural patterns
common in C/C++ are less prevalent in Rust. Once potential callees are
identified, taint is propagated in the same manner as for direct calls.


%
\System{} adopts a worklist-based algorithm~\cite{ProgAnalysis:algorithms}
adapted for inter-procedural taint analysis, as presented in
Algorithm~\ref{alg:interproc-taint}. First, it caches pointer derivations whose
sources originate from other functions, either directly (e.g., formal
parameters) or indirectly (e.g., intermediate variables derived from formal
parameters), and are therefore \emph{unresolved} within the current function
context (lines~4--10). After completing the initial pass over all functions, it
performs a depth-first search over the cached derivations to exhaustively
propagate taint (lines~11--25). This two-step process ensures that all
transitive taint relationships are resolved and that all potential temporally risky
pointers are identified. In addition, \System{} also addresses temporally risky
pointers involved with unsafe APIs, similar to the approach described in
\S\ref{sec:spa_risk}.

\begin{algorithm}[t]
\DontPrintSemicolon
\footnotesize
\KwIn{$F$ — set of functions;  $R$ — set of initially tainted raw pointers}
\KwOut{\textit{TaintedSets} — mapping from each taint source to its tainted pointers}
\SetKwBlock{Begin}{function}{end function}
\Begin(InterProcTaintPropagation{$(F,R)$})
{
    Initialize \textit{TaintedSets} to map each $r \in R$ to $\{r\}$\;
    Initialize \textit{WorkList} $\gets \emptyset$\;

    \For{each unresolved pointer derivation $D$ in $F$}
    {
        $(src,dst) \gets \texttt{ExtractSourceAndDestination}(D)$\;
        Add $D$ to \textit{WorkList}\;
        \uIf{src is tainted}
        {
            Add $dst$ to the same tainted set as $src$\;
        }
        \uElseIf{dst is tainted {\bf and} no ownership transfer in $D$}
        {
            Add $src$ to the same tainted set as $dst$\;
        }
    }

    \For{each tainted pointer $t$ in \textit{TaintedSets}}
    {
        Initialize \textit{Visited} $\gets \emptyset$\;
        Initialize \textit{Stack} $\gets \{t\}$\;
        \While{\textit{Stack} is not empty}
        {
            $p \gets$ pop an element from \textit{Stack}\;
            \uIf{$p \notin$ \textit{Visited}}
            {
                Add $p$ to \textit{Visited}\;
                \For{each unresolved pointer derivation $D$ in \textit{WorkList}}
                {
                    $(src,dst) \gets \texttt{ExtractSourceAndDestination}(D)$\;
                    \uIf{$src = p$}
                    {
                        Add $dst$ to the same tainted set as $t$\;
                        Push $dst$ onto \textit{Stack}\;
                    }
                    \uElseIf{$dst = p$ {\bf and} no ownership transfer in $D$}
                    {
                        Add $src$ to the same tainted set as $t$\;
                        Push $src$ onto \textit{Stack}\;
                    }
                }
            }
        }
    }
    \Return \textit{TaintedSets}\;
}
\caption{Inter Procedural Taint Propagation}
\label{alg:interproc-taint}
\end{algorithm}

\vspace{-2pt}
\subsection{Soundness and Precision} 
\label{sec:soundness}

\System{}'s Rust-specific static analysis is sound in identifying both
spatially and temporally risky pointers. It begins with a conservative
reachability analysis (\S\ref{sec:ana_scope}) that includes all functions that
may execute at runtime, ensuring all risky pointers that can trigger memory
errors at runtime are within the analysis scope.

To identify spatially risky pointers, \System{} begins by conservatively
annotating all instructions involving raw pointers. This over-approximation
ensures that all raw pointers are initially captured. The subsequent
analysis prunes false positives by leveraging the fact that exposed raw pointers that
bypass Rust's safety guarantees can only be used within unsafe code. Thus,
\System{} excludes 
\hl{encapsulated raw pointers that cannot be directly accessed and are protected by Rust memory safety rules.}
This pruning maintains soundness while improving precision. Additionally, \System{}
identifies and handles each unsafe API individually, based on a thorough review
of the Rust standard library.

For temporally risky pointers, \System{} employs a lifetime-aware taint
analysis. Soundness in identifying such pointers is ensured by three
key elements. 
First, the taint and pointer-derivation analyses described above are sufficient to capture all potential may-alias relationships in this context, as Rust’s ownership and borrowing rules ensure that aliasing can only occur through explicit and syntactically visible pointer derivations~\cite{RustBelt:POPL18}. 
\hl{Second, this guarantee naturally extends to loops because the LLVM IR contains a fixed and finite set of pointer derivation instructions. \System{} tracks each such instruction within a loop, analyzing only static derivation relationships rather than mutable program states (e.g., ranges, offsets, or sizes). As a result, every pointer along the derivation chain that may reference the same object as the taint source is captured, 
regardless of the complexity of the loop body or control flow.}
Third, \System{} employs a type-based analysis to conservatively resolve indirect calls, ensuring that all relevant pointer derivations are included. Together, these three elements guarantee that \System{} marks all valid alias pointers as temporally risky. 

While the analysis is sound by design, potential false negatives may arise
in practice due to implementation limitations, such as compiler optimizations or
missing IR from dynamically linked code. The approach may also introduce some
over-approximation, for example, by analyzing derivations that never occur
during actual execution. However, this imprecision is significantly reduced
compared to prior work~\cite{ERASan:Oakland24,RustSan:SEC24} relying on
traditional points-to analysis, which is unaware of pointer lifetimes. In
contrast, \System{} excludes pointers that are invalid at the time of exposed raw
pointer creation or not derived from tainted sources, as these are protected by
Rust’s compile-time safety guarantees and are not susceptible to temporal memory
safety violations.


\vspace{10pt}
\section{Lightweight Runtime Checks}
\label{sec:runtime}

In this section, we present the lightweight runtime checks of \System{}. 
\System{} uses compact memory safety metadata in place of red zones and shadow
memory to address Challenge C2 and adopts a selective instrumentation strategy
to address Challenge C3 (see \S\ref{subsec:challenge}). We describe the metadata
structures in \S\ref{sec:metadata_def} and the metadata inference approach in
\S\ref{sec:metadata_inference}. \S\ref{sec:instrumentation} details our
selective instrumentation strategy, and \S\ref{sec:runtime_check} explains how
the instrumented checks uses memory safety metadata to \hl{perform runtime validation}.

\subsection{Metadata Structure}
\label{sec:metadata_def}

For each risky pointer, \System{} maintains \emph{spatial metadata} for
spatially risky pointers and \emph{temporal metadata} for temporally risky
pointers. This metadata is inferred during static analysis, propagated at
runtime through instrumentation, and stored in dedicated data structures
rather than embedded directly in the pointer representation (e.g., fat
pointers). At runtime, the metadata is dynamically updated and used to detect
memory safety violations.

\paragraph{Spatial metadata.}
For spatially risky pointers, \System{} tracks three key attributes that are
necessary and sufficient to enforce spatial memory safety in Rust:

\begin{itemize}[leftmargin=*, topsep=1pt, itemsep=1pt] 
	\item \textbf{Capacity:} the maximum number of elements allowed in a referenced memory object.
    For pointers referencing scalar-type objects (e.g., integers), the capacity is set to 1.
 	\item \textbf{Initialized length:} the number of elements that have been initialized within a referenced memory region.
 	\item \textbf{Offset:} the index within an object that the pointer references.
\end{itemize}

\System{} maintains a map from each spatially risky pointer to its spatial
metadata, which consists of the attributes listed above; this metadata map is
stored separately from the pointers themselves.



\paragraph{Temporal metadata.}
For temporally risky pointers, \System{} tracks the following information 
as temporal metadata:
\begin{itemize}[leftmargin=*, topsep=1pt, itemsep=1pt] 
    \item \textbf{May-alias relationships:} pointers that may reference the same memory object are grouped to represent their potential aliasing.
    \item \textbf{Ownership:} the owner(s) of the referenced objects.
\end{itemize}

\System{} uses \textbf{taint source raw pointers} to link associate temporally risky pointers with their temporal metadata via two maps: a \emph{reverse map} that links each temporally risky pointer back to its originating taint source, and a \emph{forward map} that links each taint source to the corresponding metadata which it shares with its tainted pointers. This dual-mapping design avoids redundant metadata storage for multiple pointers derived from the same source while accurately associating each temporally risky pointer with its temporal metadata.

To record may-alias relationships, \System{} groups all temporally risky
pointers that may reference the same memory object into a \emph{pointer set}.
This pointer set integrates with \System{}’s taint analysis, as may-alias
pointers identified by \System{} share a common taint source and can be grouped
during taint analysis without additional effort. When an exposed raw pointer is derived
from another, their pointer sets are merged to ensure a complete representation
of may-alias relationships. 

Ownership is a critical component of temporal metadata, as the owners are
the pointers responsible for deallocating the referenced objects. As a result,
owners must be tracked at runtime to update the temporal state (i.e., dangling
or valid) of all pointers in the same pointer set. To record the owners,
\System{} maintains a dedicated \emph{owner set}, which is a subset of the
pointer set.


\subsection{Metadata Inference}
\label{sec:metadata_inference}

Metadata inference is performed during static analysis. \System{} infers each risky pointer’s memory safety metadata at its definition site and instruments the code to receive and maintain the inferred metadata, enabling runtime validation. 

\paragraph{Spatial metadata inference.} 
To infer spatial metadata for spatially risky pointers, \System{} analyzes each spatially risky
pointer’s definition site and, if necessary, traces pointer derivations back to
the allocation site of the memory object, where the spatial metadata is defined.
Specifically, \System{} backtracks along the pointer-derivation chain to locate
the corresponding \emph{root} pointer, which is the first pointer that
references the memory object. \System{} then extracts spatial metadata from the 
root pointer’s definition site, where the memory allocation appears as an 
operand. Depending on how the memory region is referenced, metadata
extraction falls into two distinct cases:

\begin{itemize}[leftmargin=*]
    \item In the \textbf{direct case}, the root pointer references a memory
    object whose spatial metadata can be directly extracted. This occurs when
    the referenced object is a basic container provided by the Rust standard
    library, such as vectors (\texttt{Vec<T>}) and arrays (\texttt{[T; N]}), 
    which manage contiguous memory
    regions. In this case, \System{} directly obtains spatial metadata from the
    container’s fields. For example, the \texttt{length} and \texttt{capacity}
    fields in \texttt{Vec<T>} directly provide \textit{Initialized Length} and
    \textit{Capacity} (\S\ref{sec:metadata_def}).
    
    \item In the \textbf{indirect case}, the root pointer refers to a memory
    object indirectly via abstractions such as \texttt{Box<T>} or
    \texttt{Rc<T>}, which do not explicitly carry spatial metadata. In this
    case, \System{} backtracks to the definition site of the underlying
    \texttt{T}-typed object to infer and extract spatial metadata.
\end{itemize}


Additionally, when the definition site of any exposed raw pointer involves pointer arithmetic, \System{} computes the resultant \textit{Offset}. 
Another category of spatially risky pointers, smart pointers associated with unsafe APIs, is relatively uncommon. Thus, \System{} handles these cases individually, applying sanitizer checks based on the semantics of each unsafe API, as discussed in \S\ref{sec:spa_risk}.

Once spatial metadata is extracted from a root pointer, it is transmitted to the
spatially risky pointers along the pointer-derivation chain. The root pointer,
along with the intermediate pointers on this chain, is referred to as
\emph{metadata-carrying pointers}. Note that metadata-carrying pointers
are not necessarily risky themselves but are tracked to enable
accurate metadata propagation. 

To enable runtime checking, \System{} propagates statically inferred spatial
metadata to spatially risky pointers through inserted instrumentation. This
process involves two cases. First, if the risky pointer is a root pointer, the
metadata is available at its definition site; therefore, instrumenting its
definition site suffices. Second, if the risky pointer is a derived pointer,
metadata must be passed along the derivation chain. In this case, \System{}
instruments the definition sites of all metadata-carrying pointers \hl{along} the chain
to ensure proper metadata propagation to the derived pointer.

\paragraph{Temporal metadata inference.}
For temporal metadata, the may-alias relationships and the taint-source raw
pointers are inferred during the identification of temporally risky pointers,
through lifetime-aware taint analysis, as described in \S\ref{sec:temp_risk}.
The owner(s) of the referenced memory objects are inferred by analyzing the
definition site of each tainted pointer based on Rust’s ownership model.
Specifically, owners are smart pointer types (e.g., \texttt{Box} and
\texttt{Rc}) that manage the lifetime of a memory object and are
responsible for its deallocation. \System{} detects owners by analyzing
pointer types and determining whether they are associated with memory
deallocation, typically indicated by their \texttt{Drop}
implementation~\cite{Drop:Rust}.



\paragraph{Robustness and completeness guarantee.} \System{} adopts a hybrid
strategy to infer spatial and temporal metadata, ensuring that metadata is
accurately and reliably extracted.

Spatial metadata is inferred for (i) exposed raw pointers and (ii) smart pointers
associated with unsafe APIs. For exposed raw pointers, \System{} extracts initial
metadata from referenced objects via static analysis and passes it to runtime
functions through instrumentation. This method is \textbf{robust} because Rust’s
ownership and borrowing rules guarantee aliasing occurs only through explicit,
visible derivations~\cite{RustBelt:POPL18}, making metadata fully traceable. It
is also \textbf{complete} since static analysis extracts only initial metadata,
while runtime instrumentation ensures precise, immediate updates. During
execution, root pointers are initialized before derived pointers, enabling
accurate metadata transmission and preventing stale values. For smart pointers
associated with unsafe APIs, \System{} addresses each case individually based on
its semantics. The small number of such APIs, combined with tailored handling, 
ensures \textbf{robustness} and \textbf{completeness}.

Temporal metadata consists of (i) may-alias relationships and (ii) object
owners. May-alias relationships are inferred through lifetime-aware taint
analysis, proven sound in \S\ref{sec:soundness}. Owners are identified based on
pointer types and \texttt{Drop} usage, following Rust’s ownership rules. As both
types and \texttt{Drop} usage are statically deterministic, this approach
guarantees both \textbf{completeness} and \textbf{robustness}.


\subsection{Selective Instrumentation}
\label{sec:instrumentation}

\System{} performs selective instrumentation by applying only the runtime checks necessary to each identified risky pointer.
To support these checks, it also instruments the program to propagate statically inferred metadata and manage it at runtime. This section introduces the five classes of instrumentation used in \System{} and how they are selectively applied according to the risky pointer types and program context.

\paragraph{Instrumentation types.}
\System{} defines five instrumentation classes (\textbf{I1}–\textbf{I5}) to initialize and update metadata during execution, and perform runtime memory safety checks based on the metadata.

\begin{itemize}[leftmargin=*, topsep=1pt, itemsep=1pt] 
    \item \textbf{I1}: Pointer activation (metadata initialization).
    \item \textbf{I2}: Spatial metadata update.
    \item \textbf{I3}: Pointer deactivation (temporal metadata update).
    \item \textbf{I4}: Spatial safety checks. 
    \item \textbf{I5}: Temporal safety checks.
\end{itemize}

\noindent\textbf{I1 type.}
\textbf{I1} instrumentation is inserted at the definition sites of \hl{identified spatially and temporally risky pointers, and metadata-carrying pointers (including root pointers).}
It \hl{receives} statically inferred metadata 
and marks each pointer as active \hl{when being triggered at runtime,} by registering \hl{the pointer} with its associated spatial \hl{or} temporal metadata.

For \textit{spatially risky pointers} and their derivation sources (i.e., root and metadata-carrying pointers), I1 establishes a runtime mapping between each pointer and its spatial metadata. Root pointers are initialized with metadata directly from static analysis, while derived pointers inherit metadata from their source pointer.
For \textit{temporally risky pointers}, I1 maintains a mapping from each taint-source raw pointer to associated temporal metadata and a reverse mapping from each tainted pointer to its taint source, as discussed in \S\ref{sec:metadata_def}. Depending on whether the definition site corresponds to a taint source or a tainted pointer, I1 creates or updates these mappings.

\paragraph{I2 type.}
\System{} employs \textbf{I2} to update spatial metadata at runtime in two scenarios: (1) when the offset of a spatially risky pointer or metadata-carrying pointer is modified (e.g., through pointer arithmetic), and (2) when the underlying memory object is modified (e.g., through container operations, such as \texttt{push()} and \texttt{pop()}). 

For (1), I2 updates the \textit{Offset} field in its associated spatial metadata. 
For (2), I2 updates and synchronizes the \textit{Initialized length} and/or \textit{Capacity} field(s) for both the pointer performing the modification and all preceding pointers in the derivation chain.
This is because those pointers all reference the same memory object. 

\paragraph{I3 type.}
\textbf{I3} instrumentation is inserted at deallocation sites, including function returns for stack-allocated objects and explicit \texttt{drop} operations for heap-allocated objects, to deactivate invalidated pointers. 
If an owner deallocates the memory object, or if the last owner is invalidated (e.g., via \texttt{mem::forget()}), I3 queries the reverse map to identify the taint-source raw pointer and updates \hl{corresponding} spatial metadata, marking the taint-source raw pointer along with all pointers in its pointer set as dangling.

\paragraph{I4 and I5 types.}
\System{} applies \textbf{I4} at pointer arithmetic and dereference sites of spatially risky pointers, and \textbf{I5} at dereference and deallocation sites of temporally risky pointers, to detect spatial and temporal memory safety violations, respectively. The detection mechanisms for I4 and I5 are detailed in \S\ref{sec:runtime_check}.

\paragraph{Instrumentation strategy.}
\System{} employs a selective instrumentation strategy that inserts only the necessary code at each instrumentation site. This approach ensures that metadata remains up-to-date and that memory safety violations are detected efficiently. Table~\ref{tab:instrumentation} summarizes the instrumentation strategy.

\begin{table}[t]
    \centering
    \scalebox{0.7} {
    \begin{tabular}{lccccc}
        \toprule
        Pointer Type & Definition & Dereference & \begin{tabular}[c]{@{}c@{}}Pointer \\ Arithmetic\end{tabular} & \begin{tabular}[c]{@{}c@{}}Container \\ Modifier\end{tabular} & Deallocation \\
        \midrule
        \begin{tabular}[l]{@{}l@{}}Spatially Risky \\ Pointers\end{tabular} & \textbf{I1} & \textbf{I4} & \textbf{I2}, \textbf{I4} & \textbf{I2} & - \\
        \hline
        \begin{tabular}[l]{@{}l@{}}Temporally Risky \\ Pointers\end{tabular} & \textbf{I1} & \textbf{I5} & - & - & \textbf{I5}, \textbf{I3}  \\
        \hline
        \begin{tabular}[l]{@{}l@{}}Metadata-Carrying \\ Pointers\end{tabular} & \textbf{I1} & - & \textbf{I2} & \textbf{I2} & - \\
        \bottomrule
    \end{tabular}
    }
    \vspace{15pt}
    \caption{\textbf{Selective instrumentation strategy.} Each class of instrumentation is applied based on the type of pointer and the type of operation, ensuring that only the necessary code is inserted at each instrumentation site. At the pointer arithmetic of a spatially risky pointer, I2 is before I4. At the deallocation site of a temporally risky pointer, I5 is before I3.}
    \label{tab:instrumentation}
\end{table}

For \emph{all three pointer types}, \System{} inserts \textbf{I1} at their definition sites to register the pointers along with their associated spatial or temporal metadata at runtime, \hl{making them activated}. This metadata is later used to initialize derived pointers or to validate memory safety at runtime.

For \emph{spatially risky pointers}, which may cause spatial memory safety violations, \System{} selectively inserts \textbf{I2} and \textbf{I4}. \textbf{I2} 
\hl{is placed}
at pointer arithmetic operations (e.g., \texttt{add()}, \texttt{offset()}) and at container modifier operations (e.g., unsafe API \texttt{set\_len()}) \hl{to update spatial metadata instantly at runtime}, 
\hl{enabling I4 to precisely perform spatial memory safety validation.}
\textbf{I4} is inserted at pointer arithmetic and dereference sites to detect spatial violations using the maintained spatial metadata. 
\hl{Importantly, I2 is placed before I4 at pointer arithmetic sites, ensuring that any invalid pointer arithmetic is detected immediately using the up-to-date metadata.}

For \emph{temporally risky pointers}, which can result in temporal memory safety violations, \System{} selectively applies \textbf{I3} and \textbf{I5}. \textbf{I3} is inserted at deallocation sites (e.g., \texttt{drop()}) to update temporal metadata, ensuring that the temporal validity state of each pointer is accurately maintained. \textbf{I5} is applied at dereference and deallocation sites to detect temporal errors, such as use-after-free and double-free. 
\hl{At deallocation sites, I5 is placed before I3 to prevent I3 from prematurely marking the pointer as invalid and causing erroneous double-free reports.}

For \emph{metadata-carrying pointers}, which serve only to transmit spatial metadata (see \S\ref{sec:metadata_inference}), \System{} selectively applies \textbf{I2} at pointer arithmetics and container modifiers (e.g., \texttt{Vec::push()}), ensuring that spatial metadata is instantly updated and accurately propagated to spatially risky pointers.
\vspace{-10pt}
\subsection{Runtime Check}
\label{sec:runtime_check}

\System{} maintains and updates memory safety metadata through I1–I3, and leverages this metadata at runtime to detect spatial and temporal violations via I4 and I5, respectively. Figure~\ref{fig:bug_pattern} (see \S\ref{sec:threat_model}) summarizes the complete set of memory errors that \System{} detects and illustrates how I4 and I5 perform runtime checks. 

For all risky pointers, \System{} first performs null checks at dereferences.
For spatially risky pointers, I4 compares the pointer’s \textit{Offset} against its \textit{Initialized Length} and \textit{Capacity}. An access is alarmed as a use-before-initialization if the offset exceeds the initialized length, or as an out-of-bounds access if it exceeds the capacity. 
For temporally risky pointers, I5 consults the temporal metadata maintained by I3 to determine whether the pointer is dangling. A dereference of a dangling pointer triggers a use-after-free alarm, while a deallocation of a dangling pointer raises a double-free alarm.

\paragraph{Benefits of our strategy.}
The benefits of \System{} are to impose lower runtime and memory overhead while providing more comprehensive detection coverage in comparison with ASan-based approaches~\cite{ERASan:Oakland24,RustSan:SEC24}.
Specifically, \System{} selectively instruments only the pointers that may potentially violate memory safety and inserts only necessary checks for them. These pointers are only a subset of the pointers that existing ASan-based techniques~\cite{ERASan:Oakland24,RustSan:SEC24} check.
Moreover, \System{} can detect memory safety bugs that existing ASan-based approaches may miss by maintaining the fine-grained spatial and temporal metadata. This metadata is compact and lightweight, contributing further to runtime efficiency.

\vspace{10pt}
\section{Implementation}
\label{sec:implementation}

We implement \System{} on top of LLVM-14.
It takes the program’s LLVM bitcode as an input, performs static analysis, applies selective instrumentation, and generates instrumented LLVM bitcode. 

The input bitcode is generated from Rust programs using a customized version of \texttt{rustc‑1.64‑nightly}. This compiler is extended to support metadata annotation during the MIR-to-LLVM IR lowering phase. Following ERASan’s~\cite{ERASan:Oakland24} annotation mechanism, we modify the \texttt{codegen-llvm} and \texttt{codegen-ssa} to insert LLVM metadata on instructions involving raw pointers. These annotations enable \System{} to identify raw pointers during static analysis, as described in \S\ref{sec:spa_risk}.
On the other hand, the output bitcode includes inserted calls to runtime functions (i.e., I1-I5 in \S\ref{sec:instrumentation}). 
At runtime, the instrumented code is invoked to update metadata instantly and detect memory safety violations as the program executes.

\vspace{10pt}
\section{Evaluation}
\label{sec:evaluation}

We evaluate \System{} in comparison with the two state-of-the-art Rust sanitizers, ERASan~\cite{ERASan:Oakland24} and RustSan~\cite{RustSan:SEC24}, as follows:
runtime overhead (\S\ref{sec:rt_overhead}), 
memory overhead (\S\ref{sec:memory}), 
compilation overhead (\S\ref{sec:compile}), and 
bug detection capability (\S\ref{sec:security}). 


\paragraph{Experiment setup.} 
All experiments were conducted on a server with an Intel Xeon Gold 6230 CPU, 80 cores, and 754 GB RAM, running Ubuntu 24.04. The benchmarks are first compiled to unoptimized LLVM IR, allowing \System{} and comparison tools to analyze program semantics and insert instrumentation before optimizations may alter or remove metadata. Each inserted check is tied to its corresponding memory operation and realized as a runtime function call, ensuring that subsequent LLVM optimizations do not eliminate it. The instrumented IR is then compiled with the default \texttt{-O3} pipeline, reflecting realistic deployment conditions. This staging preserves analysis precision while ensuring that performance measurements correspond to practical compilation, runtime, and memory costs (further explained in Appendix~\ref{sec:app:opt_issue}).

\paragraph{Benchmarks.} 
We evaluated \System{} on 28 benchmarks: 26 most frequently downloaded Rust crates from \texttt{crates.io}\footnote{\texttt{crates.io} is Rust’s official package registry. Each crate is implemented entirely in Rust and compiled as an independent unit, functioning as either a library or an executable given the benchmark input.} and two real-world applications (\texttt{servo} and \texttt{ripgrep}). 
For each benchmark, we compile and execute both the baseline versions (without instrumentation) and the instrumented versions produced by each sanitizer 20 times. The average compilation time, execution time, and memory usage are used to compute the respective overheads. For the benchmarks shared with ERASan, we use its experiment setup~\cite{web:erasan_src}. Thus, we use the same test cases to ensure a fair comparison. For the remaining benchmarks, we use their native test suites.

\paragraph{Ablation study.} 
To decompose \System{}'s overhead, we developed a variant \Variant{}, which uses \System{}’s static analysis to identify risky pointers and selectively instruments runtime checks, but employs ASan’s runtime checking instead of \System{}’s metadata-based approach. 
Comparing \Variant with \System{} isolates the benefit of lightweight metadata while comparing \Variant{} with RustSan and ERASan highlights the benefit of our precise risky pointer identification, as RustSan and ERASan use ASan's runtime checking mechanism.

\subsection{Runtime Overhead}
\label{sec:rt_overhead}

\begin{table*}[t]
\centering
\footnotesize
\resizebox{\linewidth}{!}{
\begin{tabular}{lr|rrr|rrr|rrr|rrr}
\toprule
\multirow{2}{*}{\textbf{Benchmark}} & \multirow{2}{*}{\textbf{LOC}} &
\multicolumn{3}{c|}{\textbf{Pointer Count}} &
\multicolumn{3}{c|}{\textbf{Compilation Overhead (\%)}} &
\multicolumn{3}{c|}{\textbf{Runtime Overhead (\%)}} &
\multicolumn{3}{c}{\textbf{Memory Overhead (\%)}} \\
& &
\textbf{Expo-raw} & \textbf{Risky} & \textbf{Aliased} &
\textbf{\System{}} & \textbf{ERASan} & \textbf{RustSan} &
\textbf{\System{}} & \textbf{ERASan} & \textbf{RustSan} &
\textbf{\System{}} & \textbf{ERASan} & \textbf{RustSan} \\
\midrule
base64            & 7,025     & 1,787 & 14,320 & 131,242 & 174.17 & SE     & 6,260.03 & 35.21 & -   & 431.28 & 3.56 & -    & 5,271.84 \\
byteorder         & 3,411     & 95    & 355    & 5,391   & 66.52  & 614.35 & 674.93   & 1.72  & 53.36 & 76.37  & 0.03 & 357.73 & 406.27   \\
bytes(buf)        & 5,867     & 88    & 376    & 2,904   & 47.94  & 636.77 & 748.59   & 28.39 & 137.90& 154.33 & 2.68 & 86.27  & 98.58    \\
bytes(bytes)      & 5,867     & 91    & 411    & 2,089   & 45.25  & 625.32 & 682.33   & 25.57 & 166.97& 169.62 & 2.15 & 2,218.86 & 2,143.63 \\
bytes(mut)        & 5,867     & 102   & 484    & 2,267   & 46.19  & 627.09 & 755.26   & 27.82 & 157.28& 165.48 & 2.19 & 5,376.09 & 5,339.28 \\
indexmap          & 8,693     & 386   & 2,214  & 32,132  & 103.59 & 5,807.79 & 2,310.17 & 23.06 & 287.14& 293.65 & 1.78 & 1,754.14 & 2,090.46 \\
itoa              & 613       & 9     & 32     & 291     & 47.76  & 334.31 & 414.24   & 20.34 & 116.11& 131.05 & 1.43 & 65.71  & 69.83    \\
memchr            & 1,139     & 50    & 185    & 3,133   & 86.14  & 514.86 & 560.28   & 13.18 & 212.39& 217.74 & 1.96 & 49.61  & 61.56    \\
num-integer       & 2,383     & 570   & 3,095  & 8,449   & 186.57 & 1,359.64 & 1,512.04 & 1.17  & 5.59  & 8.34   & 0.02 & 524.07 & 674.62   \\
ryu               & 3,443     & 17    & 82     & 2,247   & 64.03  & 868.87 & 931.11   & 17.71 & 63.80 & 70.89  & 0.28 & 81.69  & 85.22    \\
semver            & 2,483     & 24    & 81     & 839     & 42.34  & 390.24 & 451.72   & 4.83  & 317.21& 388.52 & 1.88 & 6,832.92 & 7,683.53 \\
smallvec          & 2,912     & 59    & 278    & 982     & 59.21  & 422.05 & 489.63   & 13.34 & 134.53& 152.33 & 1.14 & 4,370.14 & 4,518.99 \\
strsim-rs& 1,102     & 109   & 431    & 1,015   & 58.25  & 452.52 & 528.97   & 1.06  & 380.51& 389.72 & 1.36 & 5,568.87 & 5,729.60 \\
uuid(format)      & 4,971     & 15    & 50     & 62,149  & 207.82 & 9,230.63 & 2,669.96 & 40.62 & 362.41& 411.04 & 0.15 & 875.22 & 879.71   \\
uuid(parse)       & 4,971     & 15    & 50     & 62,091  & 202.32 & 9,038.49 & 2,684.37 & 37.31 & 338.06& 402.53 & 0.15 & 1,065.03 & 1,094.13 \\
\midrule
bat               & 53,517    & 2,567  & 25,546 & 138,726 & 167.91 & 25,020.17 & 4,826.05 & 321.11 & 894.97& 931.36 & 4.96 & 4,619.51 & 5,017.39 \\
crossbeam-utils   & 31,246    & 64    & 290    & 2,227   & 104.06 & 586.35 & 639.52   & 1.28  & 116.09& 136.17 & 0.05 & 57.85  & 58.97    \\
hashbrown         & 10,384    & 51    & 383    & 6,596   & 72.16  & 1,227.80 & 1,182.08 & 9.32  & 58.65 & 69.45  & 0.37 & 6,613.42 & 6,814.56 \\
hyper             & 20,952    & 1,824  & 15,201 & 114,269 & 184.98 & 20,920.41 & 5,127.40 & 43.57 & 278.16& 297.23 & 3.69 & 2,681.30 & 2,966.32 \\
rand(generators)  & 15,220    & 27    & 78     & 2,619   & 84.27  & 776.41 & 835.54   & 31.85 & 26.49 & 31.64  & 0.26 & 85.18  & 88.64    \\
rand(misc)        & 15,220    & 66    & 258    & 2,527   & 95.93  & 1,079.36 & 866.17   & 7.94  & 10.12 & 23.47  & 0.22 & 1,457.08 & 1,654.97 \\
regex             & 65,417    & 294   & 3,896  & 6,383   & 128.70 & 1,463.51 & 923.09   & 38.54 & 831.87& 867.34 & 1.83 & 8,191.68 & 8,574.25 \\
ripgrep           & 33,226    & 1,864  & 18,734 & -     & 142.07 & SEGV   & SEGV     & 304.03& -   & -    & 4.77 & -    & -      \\
syn               & 58,884    & 1,088  & 23,034 & 186,730 & 123.84 & 25,397.17 & 1,499.46 & 72.29 & 583.25& 618.92 & 1.65 & 327.11 & 343.74   \\
tokio             & 69,875    & 1,482  & 19,375 & 74,697  & 149.02 & 18,607.29 & 4,965.73 & 53.35 & 563.24& 593.22 & 1.33 & 1,343.17 & 1,504.36 \\
unicode           & 172,875   & 95    & 363    & 1,054   & 54.95  & 549.73 & 677.47   & 8.89  & 47.96 & 62.37  & 0.86 & 56.83  & 63.38    \\
url               & 40,595    & 353   & 2,266  & 34,368  & 191.89 & 1,618.63 & 1,521.94 & 21.06 & 612.75& 686.41 & 1.37 & 312.35 & 356.88   \\
servo             & 11.26 M   & 1.27 M & 14.63 M & -    & 186.13 & TO     & TO       & 86.58 & -   & -    & 2.82 & -    & -      \\
\midrule
\textbf{GeoMean}  & -       & -   & -    & -     & \textbf{97.21} & 1,635.35 & 1,193.31 & \textbf{18.84} & 152.05 & 183.50 & \textbf{0.81} & 739.27 & 861.98 \\
\bottomrule
\end{tabular}
}
\vspace{10pt}
\caption{\textbf{Pointer counts and overhead comparison.} Benchmarks are grouped by scale. Pointer counts report exposed raw pointers (Expo-raw) and risky pointers identified by \System{}, along with raw pointers plus aliases (Aliased) identified by traditional points-to analysis. Overheads are shown for \System{}, ERASan, and RustSan. Nonapplicable results are listed as -. SE indicates silent exit, SEGV indicates segmentation fault, and TO indicates a compilation timeout.}
\label{tab:overhead}
\end{table*}

The \textbf{Runtime Overhead} columns in Table~\ref{tab:overhead} show the runtime overhead introduced by \System{}, ERASan, and RustSan. Across all benchmarks, \System{} consistently exhibits the lowest runtime overhead.
By geometric mean, \System{} incurs a runtime overhead of only 18.84\%, significantly lower than ERASan's 152.05\% and RustSan's 183.50\%, presenting reductions of 87.61\% and 89.73\%, respectively.

\System{} achieves significantly lower runtime overhead for two reasons. First, it uses lifetime-aware taint analysis to precisely identify risky pointers, greatly reducing unnecessary instrumentation. In contrast, comparison tools rely on points-to analysis, conservatively treating all aliases of pointers in unsafe regions (RustSan) or raw pointers (ERASan) as risky, resulting in redundant instrumentation for pointers whose safety is already guaranteed by the Rust compiler. As shown in the \textbf{Pointer Count} columns of Table~\ref{tab:overhead}, \System{} identifies greatly fewer risky pointers than the aliased counts.
Second, \System{} employs the lightweight metadata-based runtime mechanism in place of the heavyweight ASan checks.

To quantify the contributions of these two components discussed above, we conduct an ablation study using \Variant{}, a variant of \System{} introduced earlier in this section. 
\Variant{} incurs 70.04\% runtime overhead by geometric mean, demonstrating that the use of our metadata-based checking mechanism in \System{} reduces overhead by 73.10\%. 
Compared to ERASan and RustSan, \Variant{} achieves reductions of 53.94\% and 61.83\%, highlighting the benefit of Rust-specific static analysis. We show the detailed results in Table~\ref{tab:ablation} in Appendix~\ref{sec:app:ablation}.
\subsection{Memory Overhead}
\label{sec:memory}

We measured memory overhead using the Linux \texttt{time} command~\cite{usr:bin:time}, which reports the peak resident set size (max RSS) of the process. 
This metric captures the maximum amount of memory consumption during execution, 
which is widely adopted as a realistic measure of memory overhead.

In terms of memory usage, \System{} demonstrates a substantial advantage over both comparison tools. As shown in the \textbf{Memory Overhead} columns of Table~\ref{tab:overhead}, \System{} introduces trivial memory overhead across all benchmarks, with a geometric mean of only 0.81\%. In contrast, ERASan and RustSan incur significantly higher memory overhead, reaching 739.27\% and 861.98\%, respectively.

These results highlight the contribution of \System{}’s lightweight metadata, which eliminates the substantial memory overhead imposed by shadow memory and red zone mechanisms in ASan-based tools. 
Specifically, by comparing \System{} with \Variant{} (443.90\% memory overhead), we can observe a 99.82\% reduction attributed to the metadata-based runtime checking mechanism.
Additionally, reduced instrumentation also contributes to memory savings. \Variant{} achieves 39.95\% and 48.50\% lower memory overhead than ERASan and RustSan, respectively. We describe their details in Table~\ref{tab:ablation} in Appendix~\ref{sec:app:ablation}.
\subsection{Compilation Overhead}
\label{sec:compile}

Compilation overhead refers to the additional compilation time introduced by a sanitizer's static analysis and instrumentation compared to the baseline build. As shown in the \textbf{Compilation Overhead} columns in Table~\ref{tab:overhead}, \System{} consistently incurs significantly lower overhead. By geometric mean, \System{} produces 97.21\% overhead, compared to 1,635.35\% for ERASan and 1,193.31\% for RustSan, presenting reductions of 94.06\% and 91.85\%, respectively. Both comparison tools failed to complete compilation for \texttt{servo} within a 24-hour timeout, and encountered a segmentation fault when analyzing \texttt{ripgrep} due to SVF errors.

\System{} achieves lower compilation overhead than RustSan and ERASan because both of them rely on SVF, which is a heavyweight points-to analyzer, as discussed in \S\ref{sec:bg:ptr}. 
In contrast, \System{} adopts a lightweight static analysis tailored to Rust. Despite its low cost, this analysis is sufficient to identify all spatially and temporally risky pointers, enabling selective instrumentation with modest compilation overhead.
\subsection{Security Evaluation}
\label{sec:security}

In addition to performance improvement, \System{} provides a more comprehensive memory error detection coverage than ERASan and RustSan, both of which share the same capability as ASan. 
Therefore, we show the bug detection capability of \System{} and compare it only with ASan (whose detailed approach and limitations are discussed in \S\ref{sec:bg:asan}). 


We analyzed bugs reported by RustSec---the Rust Security
Advisory Database~\cite{RustSec:2025}---over the past two years, focusing on the
cases where bug root causes (i.e., PoCs) are publicly available for validation. We
list memory safety bugs in our scope (discussed in \S\ref{sec:threat_model}) in Table~\ref{tab:bug}.
\System{} successfully
detects all of 20 bugs, whereas ASan fails to identify two out-of-bounds access
bugs, one use-before-initialization bug, and one use-after-free bug. 
These cases occur in Rust-specific contexts (illustrated as case studies in \S\ref{sec:case_spatial} and \S\ref{sec:case_temporal}). 
Because ASan was originally designed for C/C++, it effectively detects conventional memory safety violations but lacks the ability to handle Rust-specific memory safety rules and check per-pointer spatial and temporal memory safety.
In contrast, \System{} incorporates Rust’s memory safety rules in its static analysis and enforces per-pointer spatial and temporal memory safety checks, enabling the detection of such missing bugs. 
We illustrate one spatial memory safety bug in
\S\ref{sec:case_spatial} and one temporal memory safety bug in
\S\ref{sec:case_temporal}.

\begin{table}[t]
\centering
\footnotesize
\resizebox{.85\linewidth}{!}{
\begin{tabular}{lcccc}
\toprule
\textbf{RUSTSEC ID} & \textbf{Type} & \textbf{Class} & \textbf{ASan} & \textbf{\System{}} \\
\midrule
RUSTSEC-2023-0021 & NPD & Null-pointer deref & \cmark & \cmark \\
RUSTSEC-2023-0024 & NPD & Null-pointer deref & \cmark & \cmark \\
RUSTSEC-2023-0038 & OOB & Spatial & \cmark & \cmark \\
RUSTSEC-2023-0039 & OOB & Spatial & \cmark & \cmark \\
RUSTSEC-2023-0056 & OOB & Spatial & \xmark & \cmark \\
RUSTSEC-2024-0002 & OOB & Spatial & \xmark & \cmark \\
RUSTSEC-2025-0003 & OOB & Spatial & \cmark & \cmark \\
RUSTSEC-2025-0005 & OOB & Spatial & \cmark & \cmark \\
RUSTSEC-2025-0018 & OOB & Spatial & \cmark & \cmark \\
RUSTSEC-2023-0045 & UBI & Spatial & \cmark & \cmark \\
RUSTSEC-2023-0087 & UBI & Spatial & \xmark & \cmark \\
RUSTSEC-2024-0018 & UBI & Spatial & \cmark & \cmark \\
RUSTSEC-2024-0374 & UBI & Spatial & \cmark & \cmark \\
RUSTSEC-2024-0400 & UBI & Spatial & \cmark & \cmark \\
RUSTSEC-2023-0010 & DF  & Temporal & \cmark & \cmark \\
RUSTSEC-2023-0078 & UAF & Temporal & \xmark & \cmark \\
RUSTSEC-2024-0007 & UAF & Temporal & \cmark & \cmark \\
RUSTSEC-2024-0017 & UAF & Temporal & \cmark & \cmark \\
RUSTSEC-2025-0016 & UAF & Temporal & \cmark & \cmark \\
RUSTSEC-2025-0022 & UAF & Temporal & \cmark & \cmark \\
\bottomrule
\end{tabular}
}
\vspace{12pt}
\caption{\textbf{Bug detection capability of ASan and \System{}.} Listed are the 20 most recent memory safety vulnerabilities in RustSec, grouped by bug class. Additional results on older vulnerabilities are provided in Appendix~\ref{sec:bug}.}

\label{tab:bug}
\end{table}
\vspace{-10pt}


\vspace{10pt}
\subsubsection{Case Study 1}
\label{sec:case_spatial}

\sloppy
RUSTSEC-2023-0056~\cite{rustsec:2023-0056} is an out-of-bounds access vulnerability in the \texttt{vm-memory} crate~\cite{crate:vm-memory}. In this crate, \texttt{get\_slice} is a trait method intended to return a smart pointer-like abstraction, \texttt{VolatileSlice}, over a slice, but it lacks a default implementation. If a user implements this method incorrectly, for example, by miscomputing the \texttt{offset} or \texttt{count}, the internal pointer in the returned \texttt{VolatileSlice} may reference memory outside the intended region, potentially leading to out-of-bounds access.

Several methods in \texttt{VolatileMemory} trait, such as \texttt{get\_ref} and \texttt{get\_array\_ref}, invoke \texttt{get\_slice} without proper bounds checking, thereby raising potential memory safety violations. 
Listing~\ref{lst:atomic_ref} illustrates this issue using \texttt{get\_atomic\_ref} as an example. 
In line 6, \texttt{get\_slice} is invoked to wrap an allocated memory region with a requested size of \texttt{size\_of::<T>()} bytes. 
In line 9, the internal pointer of the returned \texttt{VolatileSlice} (i.e., \texttt{slice.addr}) is cast and dereferenced without verifying whether the underlying memory actually aligns with the requested bounds. If \texttt{get\_slice} returns a region smaller than the requested region, any dereference beyond the actual region results in an out-of-bounds access.

\begin{figure}[h]
\centering
\begin{lstlisting}[abovecaptionskip=1pt,belowcaptionskip=0pt,numbersep=5pt,xleftmargin=1.3em,caption={Potential OOB in \texttt{get\_atomic\_ref}.}, label={lst:atomic_ref}]
fn get_slice(&self, offset: usize, count: usize) 
                    -> Result<VolatileSlice<BS<Self::B>>>;

fn get_atomic_ref<T: AtomicInteger>(&self, offset: usize) 
                                           -> Result<&T> {
    let slice = self.get_slice(offset, size_of::<T>())?;
    slice.check_alignment(align_of::<T>())?;

    unsafe { Ok(&*(slice.addr as *const T)) }
}
\end{lstlisting}
\vspace{-10pt}
\end{figure}

According to our experiment, ASan cannot detect this bug because it only places red zones around memory objects. However, in this case, the pointer returned by \texttt{get\_slice} may point to a valid memory object, but beyond the actual valid bound, which is within this object. As a result, \textit{invalid} accesses beyond the actual bound but within the larger allocated object remain undetected by ASan, since no red zones are placed at the logical boundary returned by \texttt{get\_slice}. In contrast, \System{} tracks memory safety metadata for each pointer at its definition site. This allows \System{} to precisely extract the actual bound of \texttt{slice.addr} and perform spatial memory safety checks, detecting potential out-of-bounds access.

\vspace{-5pt}
\subsubsection{Case Study 2}
\label{sec:case_temporal}

RUSTSEC-2023-0078~\cite{rustsec:2023-0078} is a use-after-free vulnerability reported in the \texttt{tracing} crate~\cite{crate:tracing}. As shown in Listing~\ref{lst:into_inner}, the vulnerability originates from the improper use of \texttt{mem::forget} in line 4, where the exclusive owner of the underlying memory object is forgotten. While \texttt{mem::forget} prevents the object’s destructor from being called, the Rust compiler considers the object to be logically invalid after its owner is forgotten. The memory region may subsequently be reused by the compiler, making any future access to the original object via existing pointers a use-after-free violation.

\begin{figure}[h]
\centering
\begin{lstlisting}[abovecaptionskip=1pt,belowcaptionskip=0pt,numbersep=5pt,xleftmargin=1.3em,caption={Potential UAF in \texttt{Instrumented::into\_inner}.}, label={lst:into_inner}]
pub fn into_inner(self) -> T {
    let span: *const Span = &self.span;
    let inner: *const ManuallyDrop<T> = &self.inner;
    mem::forget(self);
    
    let _span = unsafe { span.read() };
    let inner = unsafe { inner.read() };
    ManuallyDrop::into_inner(inner)
}
\end{lstlisting}
\vspace{-10pt}
\end{figure}

This vulnerability stems from a violation of Rust’s ownership model rather than traditional heap misuse found in C/C++. Because the memory is never explicitly freed, ASan does not update its shadow memory to mark the region as invalid, thus fails to detect the temporal safety violations. Covering this type of vulnerability in ASan is fundamentally challenging as ASan is unaware of Rust’s ownership semantics. To detect such bugs, ASan would need to determine whether an object still has a valid owner at every program point, which requires a significant change in the underlying design of ASan.
In contrast, \System{} is designed with ownership awareness. It tracks ownership and marks the pointers referencing the same object as dangling when the last owner is dropped. Any subsequent dereferences of the dangling pointers are flagged as use-after-free. This allows \System{} to detect ownership-related memory safety violations that lie beyond ASan’s capabilities.

\vspace{0pt}
\section{Discussion}
\label{sec:limitation}

\paragraph{Type conversion bugs.}
We consider type conversion bugs, such as those introduced via unsafe APIs like \texttt{transmute()}~\cite{rust:api:transmute}, out of scope, 
as it is widely accepted as orthogonal to spatial and temporal memory safety. The same view is shared by many prior works~\cite{chen2024typepulse, CCS16TypeSan, ACSAC18Bitype}.
Type conversion bugs stem from reinterpreting one type as another, which can break safety invariants without violating spatial bounds or temporal validity. 
As a result, \System{} may not be able to detect them if they do not violate spatial bounds or temporal validity.
State-of-the-art ASan-based tools~\cite{RustSan:SEC24,ERASan:Oakland24} also share the same problem~\cite{song2019sok}.
One way to address this problem is to integrate type confusion bug detection techniques~\cite{chen2024typepulse}. But it is worth noting that \System{} is able to detect such type confusion bugs if they stem from memory errors such as UAF.

\paragraph{Cross-language attacks.}
\System{} leverages Rust’s ownership and borrowing semantics to infer memory safety metadata and enforce spatial and temporal safety. As a result, it does not guarantee the detection of memory safety violations originating from external code written in languages without such semantics, such as C/C++ libraries interfaced via FFI. Similar to ERASan and RustSan, \System{} does not cover cross-language memory safety violations, which are considered out of scope.

To address cross-language attacks, one potential direction is to integrate \System{} with existing isolation or sandboxing techniques~\cite{EternalWar:Oakland13,SanitizerSoK:Oakland19}, to mitigate memory errors originating from external code. Another direction is to extend the scope of \System{} to external libraries and enforce runtime checks at FFI boundaries.
However, this requires a deep understanding of the semantics of each external API, which is difficult to generalize and automate. It also requires static analysis on C/C++ code, which lacks Rust’s safety guarantees, making Rust-specific analysis inapplicable. Despite these challenges, exploring support for cross-language memory safety is a promising direction for future work.

\vspace{10pt}
\section{Related Work}
\label{sec:related_work}

\paragraph{Memory sanitizing.} {\System} is closely related to
ERASan~\cite{ERASan:Oakland24} and RustSan~\cite{RustSan:SEC24}, both of
which are ASan-based~\cite{ASan:ATC12} tools that 
detect memory safety errors at runtime.
These tools rely on SVF~\cite{SVF:CC16}
to identify pointers aliases with raw
pointers~\cite{ERASan:Oakland24} or used in unsafe code~\cite{RustSan:SEC24}. They retain ASan checks only
for these potentially unsafe pointers to improve performance compared to ASan.
Both tools provide the same memory safety guarantee as ASan.

In contrast, {\System} offers more comprehensive safety guarantees
while incuring drastically lower runtime and memory overhead than ERASan and RustSan.
Additionally, {\System}'s compile time significantly outperforms that
of ERASan and RustSan (\S\ref{sec:evaluation}), due to precise risky pointer identification for less instrumentation and
lightweight runtime checks.
SVF, in contrast, computes complete alias information for the whole
program, thus, is computationally expensive and unscalable for
large programs~\cite{PKRUSafe:EuroSys22}, resulting in high overhead in its client, such as ERASan and RustSan.

\paragraph{Memory isolation.}
Another major line of mitigations against unsafe code---including both
unsafe Rust source code and external C/C++ libraries---is memory isolation
for protecting safe Rust memory.
Both XRust~\cite{XRust:ICSE20} and Trust~\cite{TRust:Sec23} utilize
SVF~\cite{SVF:CC16} to identify pointer dereferences of memory used by unsafe
Rust code. XRust employs bounds checking, while Trust leverages
Intel MPK~\cite{IntelManual:2021} to enforce isolation between
memory exclusively accessed by safe Rust code and memory accessed
by unsafe code. Similarly, MetaSafe~\cite{MetaSafe:Sec24} protects
smart pointer metadata (e.g., length of {\tt String}) by storing them in a 
dedicated memory region and isolating that region using Intel MPK.
PKRU-Safe~\cite{PKRUSafe:EuroSys22} performs dynamic analysis to find unsafe 
memory accesses, motivated by concerns over SVF's performance,
and also enforces isolation via Intel MPK.
Sandcrust~\cite{Sandcrust:PLOS17} restricts external C library code
by running it in a separate process.
Fidelius Charm~\cite{FideliusCharm:CODSPY18} migrates
target sensitive data to and from protected pages before and after invoking
untrusted C libraries.

They intend to mitigate the impact of unsafe code but
allow memory errors within it. In addition to preventing
unsafe Rust code from affecting safe code, {\System} can detect memory safety
bugs within unsafe code as well as ``safe'' code. Such bugs (e.g., UAF on a
smart pointer) can arise due to issues originating from unsafe code (\S\ref{sec:temp_risk}).
Nevertheless, {\System} does not prevent directly-linked C/C++ library code
from compromising Rust programs.

\paragraph{Bug detection via static analysis.}
Static analysis is also actively explored to detect bugs in Rust programs.
Rudra~\cite{Rudra:SOSP21} identifies three common memory safety bug patterns 
and performs static analysis based on these patterns.
Similarly, MirChecker~\cite{MirChecker:CCS21} also learns existing
bug patterns and utilizes Abstract Interpretation techniques for bug detection.
Both Rupair~\cite{Rupair:ACSAC21} and SafeDrop~\cite{SafeDrop:TOSEM23}
employ data-flow analysis: Rupair addresses buffer overflows while
SafeDrop focuses on detecting invalid memory deallocations.
FFIChecker~\cite{FFIChecker:ESORICS22} also targets memory management
errors specifically caused by interactions of Rust and
external libraries through the Foreign Function Interface (FFI).
SyRust~\cite{SyRust:PLDI21}, on the other hand, uses program synthesis to 
generate test cases for testing Rust library APIs.

These tools effectively detect their respective types of bugs, but suffer from high false positive rates.
For example, Rudra~\cite{Rudra:SOSP21} reports the most bugs among
these tools but has a false positive rate as high as 89\%.
In contrast,  {\System} does not suffer from false positives---namely, each 
reported bug corresponds to a real memory safety violation, though {\System} 
may introduce redundant memory-safe checks due to the conservative
nature of static analysis.  Additionally, {\System} maintains lightweight
analysis time (\S\ref{sec:compile}), whereas static analysis tools often 
impose prohibitive analysis time for the sake of broader code coverage.
\vspace{10pt}
\section{Conclusion}
\label{sec:conclusion}

Rust provides strong memory safety through its ownership semantics and type system. However, these guarantees can be undermined by the use of unsafe code, which reintroduces memory safety vulnerabilities. To detect such bugs, ASan-based tools are commonly used. Yet, even state-of-the-art sanitizers like ERASan and RustSan incur substantial performance and memory overhead, and still fail to catch certain memory safety violations.


Therefore, we propose a novel Rust memory sanitizer, \System{}, with lower overhead and more comprehensive and accurate memory error detection than ERASan and RustSan.
We achieve this goal by precisely identifying risky pointers and selectively instrumenting those risky pointers to minimize overhead while ensuring higher detection coverage than ERASan and RustSan.
As a result, \System{} imposes 18.84\% runtime overhead, 97.21\% compilation overhead, and 0.81\% memory overhead, with geometric mean, while ERASan and RustSan, respectively, incur 152.05\% and 183.50\% runtime overhead, 1635.35\% and 1193.31\% compilation overhead, and 739.27\% and 861.98\% memory overhead. 
Furthermore, \System{} detects 55 memory safety vulnerabilities with 100\% accuracy, unlike ASan-based approaches that miss four of them.

\bibliographystyle{ACM-Reference-Format}

\clearpage
 \appendix 
 \section{Continued Bug List}
\label{sec:bug}

Table~\ref{tab:bug_old} reports the detection results of \System{} and ASan on earlier RustSec vulnerabilities, complementing Table~\ref{tab:bug}. Together, these tables cover all publicly disclosed memory safety bugs reported in the RustSec Advisory Database to date. In total, the combined dataset includes 55 vulnerabilities: 21 use-after-free (UAF), 3 double-free (DF), 21 out-of-bounds accesses (OOB), 7 use-before-initialization (UBI), and 3 null-pointer dereference (NPD). 
They also cover all the vulnerabilities experimented by ERASan.

As a result, \System{} successfully detects all the listed bugs identified by ASan, demonstrating full coverage of ASan’s detection capabilities on Rust memory safety bugs. Furthermore, as shown in Table~\ref{tab:bug}, \System{} detects additional Rust-specific bugs, including UBI and certain safe-code out-of-bounds violations. ASan fails to capture due to its reliance on coarse-grained shadow memory and red-zone mechanisms, while \System{} deploys a metadata-based solution with respect to Rust’s type model.

\begin{table}[h]
\centering
\footnotesize
\begin{tabular}{lcccc}
\toprule
\textbf{RUSTSEC ID} & \textbf{Type} & \textbf{Class} & \textbf{ASan} & \textbf{\System{}} \\
\midrule
RUSTSEC-2020-0061 & NPD & Null-pointer deref & \cmark & \cmark \\
RUSTSEC-2023-0013 & NPD & Null-pointer deref & \cmark & \cmark \\
RUSTSEC-2020-0039 & OOB & Spatial & \cmark & \cmark \\
RUSTSEC-2020-0167 & OOB & Spatial & \cmark & \cmark \\
RUSTSEC-2021-0003 & OOB & Spatial & \cmark & \cmark \\
RUSTSEC-2021-0048 & OOB & Spatial & \cmark & \cmark \\
RUSTSEC-2021-0094 & OOB & Spatial & \cmark & \cmark \\
RUSTSEC-2023-0015 & OOB & Spatial & \cmark & \cmark \\
RUSTSEC-2023-0016 & OOB & Spatial & \cmark & \cmark \\
RUSTSEC-2023-0030 & OOB & Spatial & \cmark & \cmark \\
RUSTSEC-2023-0032 & OOB & Spatial & \cmark & \cmark \\
RUSTSEC-2019-0023 & UAF & Temporal & \cmark & \cmark \\
RUSTSEC-2020-0005 & UAF & Temporal & \cmark & \cmark \\
RUSTSEC-2020-0060 & UAF & Temporal & \cmark & \cmark \\
RUSTSEC-2020-0091 & UAF & Temporal & \cmark & \cmark \\
RUSTSEC-2020-0097 & UAF & Temporal & \cmark & \cmark \\
RUSTSEC-2022-0070 & UAF & Temporal & \cmark & \cmark \\
RUSTSEC-2022-0078 & UAF & Temporal & \cmark & \cmark \\
RUSTSEC-2023-0005 & UAF & Temporal & \cmark & \cmark \\
RUSTSEC-2023-0009 & UAF & Temporal & \cmark & \cmark \\
RUSTSEC-2021-0031 & UAF & Temporal & \cmark & \cmark \\
RUSTSEC-2021-0128 & UAF & Temporal & \cmark & \cmark \\
RUSTSEC-2021-0130 & UAF & Temporal & \cmark & \cmark \\
RUSTSEC-2019-0009 & DF  & Temporal & \cmark & \cmark \\
RUSTSEC-2019-0034 & DF  & Temporal & \cmark & \cmark \\
RUSTSEC-2020-0038 & DF  & Temporal & \cmark & \cmark \\
RUSTSEC-2021-0018 & DF  & Temporal & \cmark & \cmark \\
RUSTSEC-2021-0028 & DF  & Temporal & \cmark & \cmark \\
RUSTSEC-2021-0033 & DF  & Temporal & \cmark & \cmark \\
RUSTSEC-2021-0039 & DF  & Temporal & \cmark & \cmark \\
RUSTSEC-2021-0042 & DF  & Temporal & \cmark & \cmark \\
RUSTSEC-2021-0047 & DF  & Temporal & \cmark & \cmark \\
RUSTSEC-2021-0053 & DF  & Temporal & \cmark & \cmark \\
\bottomrule
\end{tabular}
\vspace{15pt}
\caption{Detection capability of ASan and \System{} on memory safety vulnerabilities, grouped by bug class. They are memory safety vulnerabilities discovered and registered in RustSec earlier than 20 memory safety vulnerabilities in Table~\ref{tab:bug}.}
\label{tab:bug_old}
\end{table}


\section{Type (2) Unsafe APIs}
\label{sec:appen:api}

We list the Type (2) unsafe APIs that may violate memory safety, within our scope, without involving raw pointers: 
\texttt{unchecked\_add/sub/mul/neg}, \texttt{forward/backward\_unchecked}, \texttt{unchecked\_shl/shr}, and \texttt{set\_len}. 
\System{} handles them by inserting bounds or validity checking at their call sites.


\section{Ablation Study Results}
\label{sec:app:ablation}


\begin{table*}[t]
\centering
\footnotesize
\resizebox{0.9\linewidth}{!}{
\begin{tabular}{l|rr|rrr|rrr}
\toprule
\multirow{2}{*}{\textbf{Benchmark}} &
\multicolumn{2}{c|}{\textbf{Pointer Count}} &
\multicolumn{3}{c|}{\textbf{Runtime Overhead (\%)}} &
\multicolumn{3}{c}{\textbf{Memory Overhead (\%)}} \\
& \textbf{Risky} & \textbf{ASan-guarded} &
\textbf{\System{}} & \textbf{\Variant{}} & \textbf{ASan} &
\textbf{\System{}} & \textbf{\Variant{}} & \textbf{ASan} \\
\midrule
base64          & 14,320   & 1,075,072   & 35.21  & 89.27  & 624.71  & 3.56  & 3,015.06  & 10,723.03 \\
byteorder       & 355      & 24,275      & 1.72   & 17.75  & 131.23  & 0.03  & 248.62    & 1,065.10  \\
bytes(buf)      & 376      & 37,783      & 28.39  & 78.27  & 289.20  & 2.68  & 37.63     & 411.25    \\
bytes(bytes)    & 411      & 18,354      & 25.57  & 79.32  & 292.98  & 2.15  & 1,044.84  & 21,368.49 \\
bytes(mut)      & 484      & 26,796      & 27.82  & 79.64  & 295.37  & 2.19  & 3,433.34  & 64,467.47 \\
indexmap        & 2,214    & 378,711     & 23.06  & 87.86  & 419.93  & 1.78  & 1,233.98  & 4,620.00  \\
itoa            & 32       & 5,195       & 20.34  & 88.56  & 241.11  & 1.43  & 28.21     & 123.68    \\
memchr          & 185      & 16,633      & 13.18  & 53.17  & 342.19  & 1.96  & 39.85     & 81.62     \\
num-integer     & 3,095    & 75,990      & 1.17   & 5.77   & 35.12   & 0.02  & 416.44    & 769.33    \\
ryu             & 82       & 9,769       & 17.71  & 52.62  & 101.45  & 0.28  & 66.26     & 92.20     \\
semver          & 81       & 4,787       & 4.83   & 32.91  & 536.83  & 1.88  & 4,686.86  & 13,347.98 \\
smallvec        & 278      & 13,736      & 13.34  & 53.71  & 284.08  & 1.14  & 2,863.90  & 78,376.98 \\
strsim-rs       & 431      & 4,038       & 1.06   & 26.63  & 522.02  & 1.36  & 3,943.39  & 9,869.13  \\
uuid(format)    & 50       & 730,713     & 40.62  & 75.66  & 481.02  & 0.15  & 157.72    & 1,008.17  \\
uuid(parse)     & 50       & 731,856     & 37.31  & 96.46  & 467.23  & 0.15  & 166.83    & 1,239.21  \\
bat             & 25,546   & 1,859,420   & 321.11 & 542.13 & 1,187.60 & 4.96  & 1,817.64  & 36,539.25 \\
crossbeam-utils & 290      & 5,695       & 1.28   & 16.19  & 187.15  & 0.05  & 41.91     & 548.96    \\
hashbrown       & 383      & 67,106      & 9.32   & 35.51  & 124.61  & 0.37  & 4,067.88  & 28,530.65 \\
hyper           & 15,201   & 737,542     & 43.57  & 84.70  & 323.03  & 3.69  & 1,752.87  & 35,747.78 \\
rand(generators)& 78       & 17,807      & 31.85  & 55.77  & 151.86  & 0.26  & 69.77     & 378.25    \\
rand(misc)      & 258      & 9,632       & 7.94   & 30.49  & 131.42  & 0.22  & 1,050.56  & 8,025.48  \\
regex           & 3,896    & 49,383      & 38.54  & 243.58 & 1,584.07 & 1.83  & 6,739.62  & 37,082.47 \\
ripgrep         & 18,734   & 962,161     & 304.03 & 524.54 & 1,210.16 & 4.77  & 51.02     & 28,728.19 \\
syn             & 23,034   & 1,770,205   & 72.29  & 278.64 & 1,390.21 & 1.65  & 33.68     & 609.24    \\
tokio           & 19,375   & 728,064     & 53.35  & 108.46 & 914.47  & 1.33  & 769.19    & 1,843.51  \\
unicode         & 363      & 7,715       & 8.89   & 42.59  & 157.94  & 0.86  & 41.17     & 333.48    \\
url             & 2,266    & 255,893     & 21.06  & 85.74  & 937.53  & 1.37  & 206.59    & 791.42    \\
servo           & 14.63 M  & 16,994,787  & 86.58  & 218.05 & 1,281.96 & 2.82  & 8,174.46  & 52,329.43 \\
\midrule
\textbf{GeoMean}& -        & -           & \textbf{18.84} & 70.04  & 359.90  & \textbf{0.81} & 443.90 & 3,282.12 \\
\bottomrule
\end{tabular}
}
\vspace{15pt}
\caption{\textbf{Comparison of \System{}, \Variant{}, and ASan.} The table shows pointer counts (risky and ASan-guarded), runtime overhead, and memory overhead across benchmarks.}
\label{tab:ablation}
\end{table*}

The \textbf{Pointer Count} column in Table~\ref{tab:ablation} reports the number of risky pointers identified by \System{}, the total number of pointers guarded by ASan. Across all benchmarks, the number of risky pointers is substantially lower than that of ASan-guarded pointers, indicating that most pointers in Rust are guaranteed to be safe, ASan checks are excessively redundant.

Table~\ref{tab:ablation} also presents the detailed runtime and memory overhead of \Variant{}, for ablation study. These results allow for two key comparisons: (1) with \System{}, to quantify the impact of lightweight metadata-based runtime checking mechanism, and (2) with ASan-based tools, to evaluate the effectiveness of Rust-specific static analysis. Overall, both components significantly contribute to reducing runtime and memory overhead.

\vspace{10pt}
\section{Comparison of \System{} and ASan}

\System{} significantly reduces the number of instrumented pointers compared to ASan by leveraging precise Rust-specific static analysis to identify risky pointers. As shown in Table~\ref{tab:ablation}, \System{} instruments much less pointers than \System{} across all benchmarks, yielding great runtime and memory performance improvement.

Specifically, ASan incurs 359.90\% runtime overhead, while \Variant{} incurs 70.04\%, indicating that our precise risky pointer identification and selective instrumentation reduce overhead by 80.54\%. \System{} further lowers the overhead to 18.94\%, demonstrating that replacing ASan’s heavyweight shadow memory and red zones with our lightweight metadata yields an additional 73.10\% reduction. 
As for memory usage, ASan introduces 3,282.12\% overhead. \Variant{} lowers it to 443.90\%, through precise static analysis and selective instrumentation, while \System{} only presents negligible overhead of 0.81\%, owing to the lightweight metadata-based run checks.

\vspace{10pt}
\section{Benchmark Compilation Options}
\label{sec:app:opt_issue}

To ensure both accurate static analysis and evaluation on realistic production situation, we adopt a staged compilation process. For each benchmark, the compiler first emits LLVM IR with inlining and LLVM prepopulate passes disabled so that MIR-derived annotations are preserved in the IR. At this stage, \System{} and the comparison tools (ERASan and RustSan) perform static analysis and insert their respective instrumentation. After instrumentation, compilation resumes with the standard optimization pipeline to produce optimized (\texttt{i.e., -O3}) executables. 

This approach is essential rather than a shortcut. Running LLVM optimizations before analysis can replace or eliminate original instructions and drop critical metadata, leading to missed identification of risky operations. This metadata-preservation challenge is not unique to \System{} but equally affects ERASan and RustSan; analyzing directly on optimized IR would cause all these tools to miss protecting unsafe operations. By applying analysis on pre-optimized IR, we preserve full semantic information and ensure that risky pointers are precisely identified. 

At the same time, compiling instrumented IR under \texttt{-O3} guarantees that our evaluation reflects realistic deployment conditions. Most LLVM optimizations transform instructions in place rather than reordering them, so checks remain associated with the correct memory operations. Furthermore, because \System{} implements its checks as runtime calls with observable actual effects, subsequent LLVM optimizations do not remove them. Our experiments demonstrate this property in practice, as \System{} achieves $100\%$ bug detection even when executing fully optimized binaries. 

In summary, this staging is a necessary and fair methodology: it preserves metadata for accurate analysis, ensures that checks are preserved under aggressive optimization, and yields performance results under realistic deployment. Future improvements in preserving Rust-specific metadata across optimization could streamline this process, but the present design is the only viable way to ensure correctness across Rust memory safety sanitizers.


\end{document}